%% file: main.tex
\pdfoutput=1 
\newif\ifarxiv
\arxivtrue

\documentclass[journal]{IEEEtran}

\usepackage[utf8]{inputenc}   
\usepackage[T1]{fontenc}
\usepackage{booktabs}
\usepackage{graphicx}
\usepackage{amsmath,amssymb}
\usepackage{CJKutf8}          
\usepackage[hidelinks]{hyperref}
\usepackage{url}
\usepackage{bm}
\usepackage{orcidlink}
\usepackage{makecell}
\usepackage{multirow}

\newcommand{\jp}[1]{\begin{CJK}{UTF8}{ipxm}#1\end{CJK}}

\ifarxiv
  \usepackage[absolute]{textpos}
  \newcommand{\arxivnotice}{%
    \begin{textblock*}{\textwidth}(0.5in,0.35in)
      \centering\footnotesize This work has been submitted to the IEEE for possible
      publication. Copyright may be transferred without notice, after which this
      version may no longer be accessible.
    \end{textblock*}}
\else
  \newcommand{\arxivnotice}{}
\fi

\begin{document}

\title{A Geometry-Limited Identification Floor and Its Consequences for
Voice-Clone Attribution in Professional Voice Actors}

\author{Shuhei~Kato\,\orcidlink{0000-0001-5746-6811}%
\thanks{Corresponding author: Shuhei Kato (e-mail: shuhei@shuheikato.info).}%
\thanks{The author is an independent researcher in Tokyo, Japan.}}

\maketitle
\arxivnotice

\begin{abstract}
\input{abstract}
\end{abstract}


\input{body}

\input{figures}

\input{acknowledgment}

\bibliographystyle{IEEEtran}
\bibliography{refs}

\end{document}

%% file: abstract.tex
A voice actor's voice is their asset, and AI cloning directly threatens it. The natural defense flags the enrolled actor whose embedding similarity to a suspect recording crosses a threshold. We show it fails where it is most needed: trained voices crowd the embedding space, and each actor performs many styles. On 1,168 Japanese voice actors (56,568 segments, \textasciitilde63\,h), a misidentification floor survives calibration, score normalization, and discriminative re-ranking (linear and nonlinear, including PLDA): the residual is a limit of the embedding geometry, not of the back-ends we evaluate. The best ensemble still leaves \textasciitilde2.6\,\% closed-set misidentification, several-fold above matched controls; session-disjoint, re-ranking lowers the floor only to 13.0\,\%. The same crowding drives false attribution: on a generic English encoder, roughly half the clones of non-enrolled people falsely accuse an enrolled actor, while --- by a separate real-vs-synthetic shift --- 32\,\% of Seed-VC clones of enrolled targets are missed at the same threshold; one operating point couples the two, and none escapes both. A domain-matched, voice-actor-trained encoder mitigates substantially (a four-fold gender gap vanishes; wrongful misattribution falls to 1.5--10\,\%), but does not remove the floor. Controls (codec, channel, vocoder, content) support reading the miss rate as a real-versus-synthetic covariate shift, not missing speaker information. Fixed-threshold clone attribution is thus unreliable here, and on a generic encoder unfair. Robust attribution must extend spoofing-aware speaker verification to open-set $\ifarxiv\bm{1{:}N}\else 1{:}N\fi$ (anti-spoofing gate, domain-matched encoder, per-speaker calibration, abstain option), and even then supports detection, not autonomous enforcement.

%% file: body.tex
\section{Introduction}
\label{sec:introduction}

\begin{figure*}[!t]
\centering
\includegraphics[width=\textwidth]{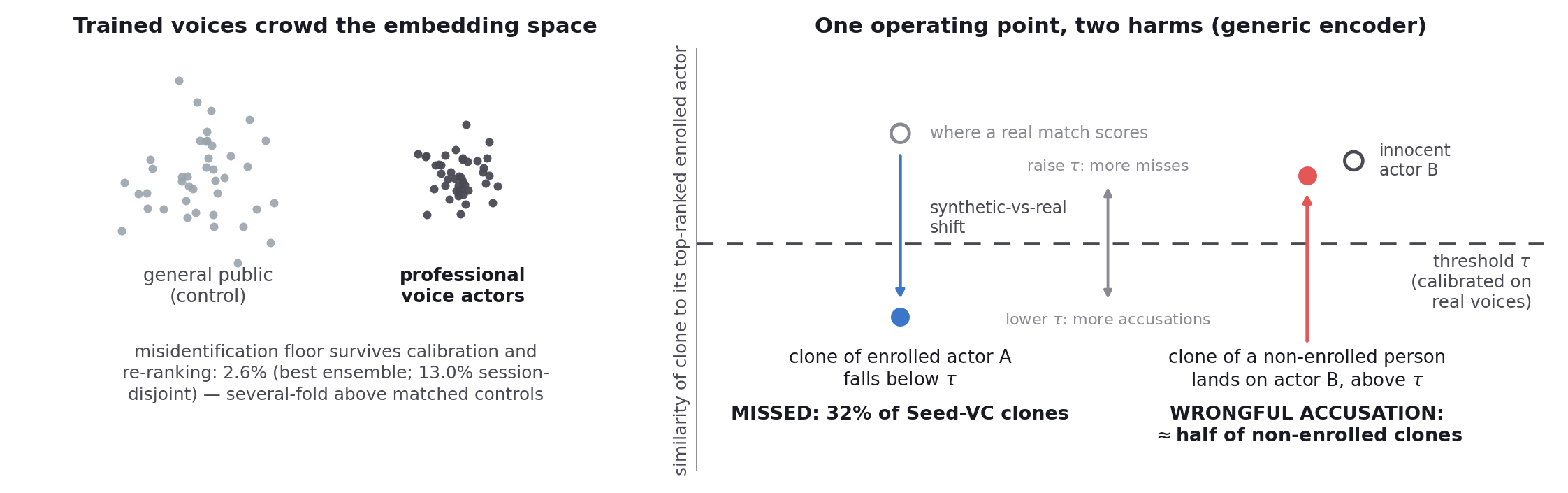}
\caption{The failure this paper quantifies. In the dense embedding space of professional voice actors (left), a similarity-threshold detector fails both ways at one operating point (right): a clone of an \emph{enrolled} actor is pushed below the real-calibrated threshold $\tau$ by the synthetic-vs-real shift (missed), while a clone of a \emph{non-enrolled} person lands on an innocent enrolled actor above $\tau$ (wrongful accusation). Raising $\tau$ trades one harm for the other; the limit is the embedding geometry, not the threshold. Rates are for the generic English encoder (Section~\ref{subsec:defensive_clone_probe_and_clone_geometry}).}
\label{fig:overview}
\end{figure*}

AI voice cloning generates a target speaker from seconds of audio \cite{wang2023valle}. For a professional voice actor, whose voice is their livelihood, this is an existential threat~\cite{almeda2025labor} --- one the profession itself has organized against, most visibly in the ``NOMORE\jp{無断生成AI}'' (No More Unauthorized Generative AI) campaign launched in 2024 by 26 professional Japanese voice actors \cite{nomore2024}. The natural technology-based defense is to make the machine tell voices apart: embed a suspect utterance, compare it to enrolled actors, and flag actor \emph{X} when the similarity to \emph{X} crosses a threshold --- reducing ``Is this an unauthorized clone of \emph{X}?'' to a $1:N$ decision. This paper shows that the machine is worst at exactly the voices of this profession (Fig.~\ref{fig:overview}).

\textbf{Two scenarios make the stakes concrete.}
\emph{(1) Wrongful accusation.} A non-actor clones their own (or a friend's) voice; it happens to resemble some enrolled actor \emph{B}, and an objective-metric detector points at \emph{B}. As of the U.S. Copyright Office's July 2024 review \cite{usco2024digitalreplicas}, no U.S. federal statute directly addresses unauthorized voice replication and the main remedies are civil; at the state level, Tennessee's ELVIS Act (2024) \cite{tennessee2024elvis} extended that state's existing civil-and-misdemeanor right of publicity to voice and to tools whose primary purpose is voice cloning --- liability attaching to unauthorized \emph{use and distribution} rather than to cloning as such. In Japan, where the actors we study work, no statute addresses it either: the publicity right is case-law only (\emph{Pink Lady v. Kobunsha}, Sup.\ Ct., Feb.\ 2, 2012~\cite{pinklady2012}, framed around the customer-drawing power of a persona) and its application to voice is untested in court. Either way, a civil claim resting on this metric would be damaging. A sharper variant: even a company that \textbf{legitimately licenses actor \emph{A} and clones \emph{A} with consent} can be sued by an unrelated actor \emph{B} the clone happens to resemble, the metric naming \emph{B} would be damaging evidence. \emph{(2) Unprotected victim.} Conversely, an actor's voice \emph{is} cloned without permission, but the synthesizer is mediocre: the result \textbf{audibly sounds like the victim to a human ear, yet its objective similarity falls below the threshold}, so the victim cannot be protected. (The human ear is the standard such a claim appeals to, not ground truth --- Section~\ref{sec:related_work} \cite{hayashi2026vtuberfans}.) Loosening the threshold to fix this then floods scenario (1) with false accusations. These failures are not independent bugs: they are \textbf{the two sides of one operating point (threshold)} --- the false-accept/false-reject trade-off. Worse, falling below the threshold is not proof of non-identity either --- by the real-synthetic shift of Section~\ref{subsec:defensive_clone_probe_and_clone_geometry}, even a clone derived from the person themselves can fall below it --- so the metric hands a decisive claim to neither side. And because the error's source lies in the embedding-space geometry, \textbf{adjusting the threshold alone cannot resolve both at once}. This is the danger of fixed-threshold attribution that the rest of the paper quantifies. No court or platform runs this exact procedure today; we quantify the failure mode any naive similarity-threshold system would inherit, and show that thresholding alone cannot make it safe.

Our contribution is to show what these effects produce when they compound in one domain: where both the \textbf{inter-speaker density} (how close the trained voices are) and the \textbf{domain-intrinsic intra-speaker style range} (narration, dialogue, character voices) are high, the \textbf{rank-1 identification error --- closed-set identification that names the query's nearest-neighbor speaker (Section~\ref{subsec:raw_geometry}) --- is limited by the embedding geometry, not by scoring}, with concrete, partly \textbf{mitigable} consequences for clone attribution and fairness. The ingredients are individually known; their joint effect in the high-density professional voice-actor domain is not. That a raw cosine threshold is suboptimal has been clear since the era of i-vectors and probabilistic linear discriminant analysis (PLDA) \cite{prince2007plda,najim2011ivector,kenny2010bayesian,garciaromero2011analysis} --- the reason adaptive symmetric normalization (AS-norm) \cite{cumani2011asnorm} and calibration \cite{brummer2006cllr} exist --- as have hubness in high-dimensional embeddings \cite{radovanovic2010hubs} and the fact that test-side score normalization such as T-norm \cite{auckenthaler2000score} is a \emph{monotone} transform of each query's scores, moving the threshold but not the neighbor ranking \cite{cumani2011asnorm}. AS-norm additionally carries an enrollment-side term whose cohort statistics differ across speakers, so it \emph{can} re-rank neighbors --- empirically, only slightly here. The forensic-voice-comparison literature has likewise long studied how the choice and density of the relevant population affect likelihood-ratio evidence \cite{rose2002forensic,morrison2009paradigm,hughes2015population}. Because no \emph{monotone} recalibration can move a rank-1/argmax decision, the load-bearing test is not calibration but \textbf{discriminative re-ranking that genuinely reorders neighbors}: linear discriminant analysis (LDA) \cite{fisher1936use}, within-class covariance normalization (WCCN) \cite{hatch2006within}, and a two-covariance PLDA log-likelihood ratio (LLR) \cite{kenny2010bayesian,garciaromero2011analysis,brummer2010speaker}, fit by moments and cross-checked by an expectation-maximization (EM) refit (Appendix~A). These only \emph{partly} reduce the residual misidentification floor and hub structure; the remainder is a property of the geometry itself.

\textbf{Contributions.}

\begin{enumerate}
\def\labelenumi{\arabic{enumi}.}

\item
  A provenance-tracked, reproducible corpus and pipeline for voice-actor embedding geometry, with a documented coverage funnel (Section~\ref{sec:data_and_pipeline}).
\item
  Across eight encoders and two ensembles, a closed-set rank-1 identification analysis showing a \textbf{misidentification floor that survives discriminative re-ranking (LDA, WCCN, and a two-covariance PLDA LLR --- moment-fit and EM alike), not merely calibration}, together with consistently positive but multiple-comparison-fragile $k$-occurrence hubness (Benjamini--Hochberg-significant, Holm-nonsignificant; Sections~\ref{subsec:raw_geometry} and~\ref{subsec:misidentification_floor_survives}).
\item
  Confound control (codec, recording channel, agency studio) and matched comparison to control populations (JVS \cite{takamichi2019jvs}, Common Voice JA \cite{ardila2020commonvoice}), showing the domain is genuinely harder, driven partly by inter-speaker proximity and partly by intra-speaker style range (their relative shares not fully separated; Section~\ref{subsec:domain_is_genuinly_high_density}).
\item
  A defensive clone probe quantifying missed attribution (an attribution-recall miss, not a countermeasure false-negative) and wrongful accusation at deployment scale, with a clone-geometry analysis attributing the failure to a \textbf{synthetic-vs-real covariate shift}, not lack of identity information (Section~\ref{subsec:defensive_clone_probe_and_clone_geometry}).
\item
  Five cross-cutting findings (A--E): the synthetic-real shift (A), a \textbf{gender fairness gap mitigable by a domain-matched encoder} (B), style-dependent misattribution risk (C), the cost of a single operating point (D), and the community/reciprocity structure of confusions (E) --- each worse for the generic English encoder and mitigated by a Japanese voice-actor-trained one (Sections~\ref{subsec:defensive_clone_probe_and_clone_geometry} and~\ref{subsec:cross_cutting_findings}).
\end{enumerate}

Together, these yield \textbf{concrete design requirements} for anyone deploying voice-clone attribution or training-data-misuse detection --- a spoofing-aware, open-set attribution pipeline with a domain-matched encoder, per-speaker calibration, and an abstain-and-human-review option (Section~\ref{sec:discussion}).

\section{Related work}
\label{sec:related_work}

\textbf{Speaker embeddings and verification.} The standard encoders based on the time-delay neural network (TDNN)  \cite{waibel1989phoneme} that extract a fixed-length speaker-identity vector from an utterance are the x-vector \cite{snyder2018xvector} and ECAPA-TDNN \cite{desplanques2020ecapa}. More recently, self-supervised models (WavLM \cite{chen2022wavlm}, HuBERT \cite{hsu2021hubert}) and newer architectures (CAM++ \cite{wang2023campp}, ReDimNet \cite{yakovlev2024redimnet}, and the related ERes2Net \cite{chen2023eres2net}) have driven the equal error rate (EER) on VoxCeleb \cite{nagrani2017voxceleb} below 1\,\%. In deployment, raw cosine similarity is not used directly for a threshold decision; instead, a PLDA back-end \cite{prince2007plda,kenny2010bayesian,garciaromero2011analysis}, AS-norm \cite{cumani2011asnorm,matejka2017norm}, and calibration \cite{brummer2006cllr} make thresholds transferable across recording conditions. AS-norm normalizes each query's score relative to its similarity distribution within a cohort of unrelated speakers; we evaluate it under this normalization. The discriminative core of PLDA --- a two-covariance PLDA LLR --- is one of the neighbor-reordering re-rankers in our suite (Section~\ref{subsec:misidentification_floor_survives}). The main tables use a moment/scatter fit, whose size-weighted between-class scatter is a biased (over-)estimate (Appendix~A). To rule the estimator out as the bottleneck, we also refit the same model by EM/maximum-likelihood (ML) --- the canonical construction for the length-normalized TDNN encoders (x-vector, ECAPA-TDNN, CAM++, ReDimNet) --- on the same splits. Closed-set misidentification moves by at most 0.4 percentage points (pp) on every properly-configured encoder (Appendix~A), so the residual floor is not an artifact of the moment fit. A \emph{full} generative PLDA back-end over the self-supervised front-ends (WavLM, JP-HuBERT) remains future work; those are diagnostic probes in any case (Section~\ref{subsec:raw_geometry}).

\textbf{Hubness.} In high-dimensional nearest-neighbor spaces, a few points become the nearest neighbor of disproportionately many others --- \emph{hubs} \cite{radovanovic2010hubs}. Hubness is conventionally measured by the skewness of the $k$-occurrence distribution (how many times each point is another's nearest neighbor), and reduced by centering, mutual proximity \cite{schnitzer2012mp}, non-iterative contextual dissimilarity measure (NICDM) \cite{jegou2010accurate}, or cross-domain similarity local scaling (CSLS) \cite{lample2018word} --- our ``centering helps WavLM'' is one example. Our main results use centering and AS-norm, but we also ran a systematic comparison of these methods (cf.~\cite{feldbauer2019hubness}): on the same query$\rightarrow$centroid rank-1 protocol, comparing raw cosine, centering, CSLS ($k$=10), empirical mutual proximity \cite{schnitzer2012mp}, and NICDM ($k$=10) across four encoders (ECAPA-TDNN, animeva, WavLM, SV-4; animeva is our domain-matched Japanese voice-actor ECAPA-TDNN, defined in Section~\ref{sec:data_and_pipeline}), no method improves misidentification by more than \textasciitilde0.3 points over raw cosine (local scaling \emph{hurts}, by up to 5.8 points), and even mutual proximity --- which flattens the hub structure best (e.g. animeva $k$-occurrence skewness 0.75$\rightarrow$0.28) --- leaves misidentification essentially unchanged. Hubness can be reduced while the error floor does not move: direct evidence that the floor is a property of the geometry, not an artifact of hubness. Our claim is that a floor remains \emph{after} these standard reductions, not that the reductions are ineffective.

\textbf{Voice privacy/anonymization.} Our defensive setting is adjacent to speaker anonymization and voice privacy work \cite{tomashenko2020voiceprivacy}, which deliberately \emph{suppresses} speaker identity. We measure the converse: how reliably can identity be \emph{attributed} to a synthetic clone?

\textbf{Anti-spoofing and spoofing-aware speaker verification (SASV).} ASVspoof \cite{wu2015asvspoof,yamagishi2021asvspoof} addresses detection of synthetic and spoofed speech, with tandem CM--ASV operation assessed by the tandem detection cost function (t-DCF) \cite{kinnunen2018tdcf}; SASV \cite{jung2022sasv} integrates these countermeasures with speaker verification, and the latest edition, ASVspoof~5 \cite{wang2024asvspoof5}, folds SASV evaluation into the challenge at crowdsourced scale. The closest existing resource to our threat setting is SpoofCeleb \cite{jung2025spoofceleb}, which trains 23 text-to-speech (TTS) systems on in-the-wild speech of 1,251 real VoxCeleb1 speakers and benchmarks deepfake detection and SASV against that real enrolled population --- still as a per-trial binary decision. A known weakness: countermeasures generalize poorly to unseen synthesis methods \cite{muller2022generalize}. Our task --- attributing a clone to a real speaker's identity in an open set --- extends SASV's target/non-target-plus-spoof binary decision to $1:N$-way attribution. A two-party dispute in which the suspected target actor is already named may reduce to $1:1$ verification; even then, the same real-synthetic shift, calibration-transfer, and per-speaker threshold-spread problems apply (Sections~\ref{subsec:misidentification_floor_survives} and~\ref{subsec:defensive_clone_probe_and_clone_geometry}). We adopt the more general $1:N$ framing, subsuming the screening use case --- matching audio of unknown origin against a roster --- that motivates this paper. This $1:N$ direction distinguishes our study from the known vulnerability of $1:1$ verification to cloned audio (an impostor clone falsely accepted against its \emph{claimed} target): our object is not threshold placement but which of $N$ real actors a clone is attributed to, and a floor on that decision that survives discriminative re-ranking. Existing large-gallery open-set identification benchmarks~\cite{lin2024voxblink2} target the general population; our contribution is that the difficulty and the attribution floor are driven by the domain-intrinsic inter-speaker proximity and intra-speaker style range of the trained professional voices, which are largely absent in such corpora.

\textbf{Deepfake source attribution.} A separate line of work attributes generated audio to the model that produced it (vocoder fingerprints~\cite{yan2022vocoderfp}, source tracing~\cite{klein2024sourcetracing}), and the recent source-verification wave asks whether two synthetic tracks share a \emph{generator} (SV-inspired but generator-level)~\cite{negroni2025source}. All of this is \emph{generator}/model attribution; at the speaker level, Cai et al.~\cite{cai2023sourcespeaker} show the \emph{source} speaker behind a voice-conversion attack remains identifiable from the converted audio. Our question is orthogonal to both: which \emph{target} real actor a single clone is attributed to, over a roster of real speakers. We do not evaluate source-attribution methods; as a side observation, Section~\ref{subsec:defensive_clone_probe_and_clone_geometry} reports that in a \emph{speaker}-embedding space, clustering by synthesizer is weak relative to clustering by speaker --- we make no claim about what dedicated source-attribution methods can achieve.

\textbf{Fairness in speaker recognition.} Demographic disparities in speaker recognition are documented \cite{hutiri2022bias}. We show a gender disparity consistent with originating in the encoder's training domain and language, and removed by a domain-matched model.

\textbf{Japanese/voice-actor speech.} For controls and a domain-matched encoder, we use JVS \cite{takamichi2019jvs}, Common Voice \cite{ardila2020commonvoice}, JTubeSpeech \cite{takamichi2021jtubespeech}, and animeva \cite{animeva}, an ECAPA-TDNN trained on Japanese voice-actor speech (``animeva'' is our shorthand, defined in Section~\ref{sec:data_and_pipeline}). For clone generation, we use the zero-shot voice-conversion model Seed-VC \cite{liu2024seedvc} (v2 release; a conditional-flow-matching (CFM) path with an autoregressive component, official default AR/CFM checkpoints, 30 sampling steps), the zero-shot TTS system GPT-SoVITS~\cite{gptsovits} (run reference-only, no per-speaker fine-tuning), which spans \textbf{two architecture lines} --- an incremental v1/v2 line with a high-fidelity \texttt{v2ProPlus} refinement, and a newer reference-faithful v3/v4 line --- each run with its official pretrained checkpoint pair (v1: \texttt{s2G488k.pth}, v2: \texttt{s2G2333k.pth}, v2ProPlus: \texttt{s2Gv2ProPlus.pth}, v3: \texttt{s2Gv3.pth}, v4: \texttt{s2Gv4.pth}, with the corresponding default text-to-semantic (T2S) models), and the zero-shot TTS system Irodori-TTS~\cite{irodoritts} (Irodori-TTS-500M-v3, with the Semantic-DACVAE-Japanese-32dim codec~\cite{dacvae_ja}) --- the latter two are public systems without academic papers, selected as representative Japanese-capable zero-shot synthesis; their fidelity is verified internally via the speaker-similarity measurements in Section~\ref{subsec:defensive_clone_probe_and_clone_geometry}, and because these systems evolve rapidly, exact model IDs and commits are released with the provenance records of Section~\ref{sec:ethics_and_reproducibility}.

\textbf{Acting voices and human identity judgments.} Yamamoto et al.~\cite{yamamoto2026actingsv} evaluate a VoxCeleb-trained ECAPA-TDNN --- our generic baseline --- on ten female voice actors performing five character styles over identical text \cite{hayashi2026jasj}: EER is 2.60\,\% between same-style utterances but 25.10\,\% across styles, against 0.81\,\% on VoxCeleb1. Their axis is intra-speaker (style collapses genuine scores), ours inter-speaker (density collapses impostor margins over 1,168 identities): the two error directions of the trade-off of Fig.~\ref{fig:overview}, measured concurrently by an independent group. Hayashi et al.~\cite{hayashi2026vtuberfans} measure the corresponding human decision: habitual viewers of a VTuber ($n = 798$) judged same/different speaker on acting-voice pairs matched in ECAPA-TDNN cosine across the familiarity contrast ($\approx$0.47 same-speaker, $\approx$0.18 different-speaker). On pairs where neither voice was familiar, the same listeners showed \emph{negative} sensitivity ($d' = -0.71$); on pairs including the voice they follow, weakly positive sensitivity ($d' = 0.46$) with a liberal criterion ($C = -0.38$), 56\,\% accepting a different-speaker pair that our encoders separate comfortably. The design is preliminary (one target, one trial per condition, fixed order with author-acknowledged expectation effects), and its direction matches earlier laboratory evidence that identity judgments degrade on acting voices \cite{hayashi2019bulletin}; what carries is that the human ear is no clean reference --- both instruments fail here, not always in the same direction --- and that familiarity moves sensitivity and criterion alike, so ``listeners would identify $X$'' is underdetermined until the listener population is specified.

\section{Data and pipeline}
\label{sec:data_and_pipeline}

\textbf{Registry (Phase 0a).} We archive the \url{seigura.com} voice-actor directory~\cite{seigura_directory} (\jp{声優名鑑}, the \emph{Seiyu Grand Prix} directory; operator Imagica Infos Co., Ltd.; URL and access dates in the references and provenance log). It lists \textbf{1,839 speaker entries across 255 agencies} (1,838 distinct normalized names --- one name maps to two entries; Section~\ref{sec:limitations_and_future_work} dedup), with each actor's name, agency, and credited works, but no audio. This registry is the denominator for our coverage. Every fetch is logged with its provenance (URL, UTC timestamp, SHA-256).

\textbf{Audio (Phase 0b).} We collect agency audio through per-agency adapters --- each agency's site exposes samples differently, so each needs its own scraper: we enumerate the talent roster, download each talent's published samples (multiple speaking styles per talent), and match them back to registry actors by normalized name. Adapters exist for the 45 highest-actor-count agencies --- \textasciitilde18\,\% of the 255 registry agencies but \textasciitilde64\,\% of registry actors, because agency size is heavily skewed; the remaining \textasciitilde82\,\% are a long tail of a few actors each, left un-adapted at the manual per-site cost of a new adapter (hence a selection bias toward larger agencies, Section~\ref{sec:limitations_and_future_work}). For freelance actors we use identity-verified official personal sites and official YouTube channels. ``Agency-affiliated'' means having a registry agency field other than ``freelance'' (a one-person agency counts). This yields \textbf{1,103 agency-affiliated actors (7,128 files) plus 71 freelance actors (277 files)} --- 1,174 actors in total.

\textbf{Preprocessing (Phase 0c).} We standardize everything to 16\,kHz mono, remove silence with energy-based voice-activity detection (VAD; librosa \cite{mcfee2015librosa} \texttt{effects.split}, threshold 30\,dB below the reference level), and split into consecutive fixed 4\,s windows, retaining a trailing remainder only if $\geq$3\,s (so segments span 3--4\,s) --- the cut boundaries are mechanical, unrelated to sentence boundaries; speaker embeddings capture identity content-independently, and the same segmentation is applied to every encoder and control population, so comparisons remain fair. 24 source files failed decoding or were emptied by VAD, and speakers left with zero usable segments were dropped, giving a final corpus of \textbf{56,568 segments / 1,168 speakers (1,103 agency + 65 freelance) / \textasciitilde63\,h}. Each segment carries a provenance back-pointer to its source and a \texttt{recording\_source} label. Coverage against the registry is \textbf{1,168/1,839 = 63.5\,\%}. The 671 uncollected actors are mostly affiliated with smaller agencies for which no scraper adapter was built (511 actors --- we built adapters for 45 of the registry's 255 agencies), plus freelance sources that could not be captured (e.g. embedded players); this agency-side selection bias is discussed under generalization limits in Section~\ref{sec:limitations_and_future_work}.

\textbf{Encoders.} The six base encoders are: x-vector~\cite{snyder2018xvector} and ECAPA-TDNN~\cite{desplanques2020ecapa} (both English / VoxCeleb); WavLM-base-plus-sv~\cite{chen2022wavlm} (English self-supervised model with a verification head); CAM++~\cite{wang2023campp} (Chinese; ModelScope checkpoint trained on \textasciitilde200k Mandarin speakers); ReDimNet-b2~\cite{yakovlev2024redimnet} (English); and a layer-mean of Japanese HuBERT~\cite{reazon_jhubert} (a Japanese self-supervised probe trained on the ReazonSpeech TV-broadcast corpus~\cite{yin2023reazonspeech}). For language/domain controls, we add two Japanese speaker verification models, referred to throughout by short names of our own: \textbf{jxvector} (the \texttt{xvector\_jtubespeech} model, trained on JTubeSpeech) \cite{jxvector} and \textbf{animeva} (the \texttt{anime\_speaker\_embedding} model, an ECAPA-TDNN trained on \textasciitilde989 Japanese voice actors) \cite{animeva}; an overlap audit by name-matching against its training set's published Visual Novel Database (VNDB)~\cite{vndb} voice-actor metadata (Section~\ref{subsec:cross_cutting_findings}) finds 21\,\% of our corpus in its training set, but its advantage holds on the held-out 79\,\%. animeva shares the ECAPA-TDNN architecture; throughout, ``ECAPA-TDNN'' unqualified denotes the generic English/VoxCeleb model, and the domain-matched model is always ``animeva''. The two ensembles use \textbf{embedding-level fusion}: we concatenate each model's L2-normalized embedding and renormalize, so the cosine similarity of the concatenated vector equals the mean of the per-model cosine similarities (SV-4 bundles the four verification models; all-6 bundles all six). This is deliberately a \textbf{simple mean-pooling baseline}, not a novel fusion method --- one of the mitigations we test. Stronger fusion does not lower the floor, so we keep the simple fixed-weight mean-pool (Table~\ref{tb:robustness}).

\textbf{Legal/ethics.} Collection and analysis were conducted entirely within Japan under Article 30-4 (information analysis) of the Japanese Copyright Act~\cite{japan_copyright_act}, subject to that Article's proviso, which we assess in Section~\ref{sec:ethics_and_reproducibility}; no audio is redistributed, and the clones are used only for the measurements in this paper (no distribution, service offering, or impersonation; Section~\ref{sec:ethics_and_reproducibility}, including other jurisdictions).

\section{Results}

\subsection{Raw geometry: high accuracy, thin margins}
\label{subsec:raw_geometry}

The strong encoders look excellent, which is exactly the problem: high rank-1 accuracy coexists with a large mass of queries decided by a hair. Table~\ref{tb:raw_geometry} reports raw-cosine geometry across all eight encoders and the two embedding-level ensembles. Rank-1 accuracy --- the fraction of queries whose nearest neighbor is the correct speaker --- is high for the strong encoders (CAM++ and ReDimNet at 92.6\,\%, ensembles at \textasciitilde95\,\%, domain-matched animeva at 97.3\,\%), and every encoder exhibits positive $k$-occurrence hubness skew.

\begin{table*}[!t]
\centering
\caption{Raw-cosine geometry per encoder and ensemble (no normalization): rank-1 identification accuracy, misidentification rate (margin \textless{} 0), and $k$-occurrence hubness skew. This is a \emph{closed-set} setting (every query's true speaker is enrolled, no reject option); rank-1 counts strictly-positive-margin queries and misID strictly-negative, so misID = 1 \(-\) rank-1 \emph{up to zero-margin ties} --- which fall in neither and account for the small $\leq$0.2 percentage-point (pp) shortfall in the ensemble rows. The two columns are complementary views of the same quantity, shown together for readability. The reject-capable operating metrics --- still mated, i.e. not non-target-tested --- are the $1:N$ EER of Section~\ref{subsec:misidentification_floor_survives} and the threshold-conditioned clone attribution of Section~\ref{subsec:defensive_clone_probe_and_clone_geometry}.}
\label{tb:raw_geometry}
\begin{tabular}{lrrr}
\toprule
Encoder & Rank-1 & misID & Hubness skew \\
\midrule
x-vector (EN/VoxCeleb) & 80.0\,\% & 20.0\,\% & 3.6 \\
ECAPA-TDNN (EN/VoxCeleb) & 86.7\,\% & 13.3\,\% & 2.5 \\
WavLM-base-plus-sv (EN) & 27.3\,\% & 72.7\,\% & 1.4 \\
CAM++ (ZH) & 92.6\,\% & 7.4\,\% & 2.2 \\
ReDimNet-b2 (EN) & 92.6\,\% & 7.4\,\% & 2.2 \\
JP-HuBERT (layer-mean) & 65.9\,\% & 34.1\,\% & 1.9 \\
jxvector (JTubeSpeech) & 69.4\,\% & 30.6\,\% & 1.5 \\
animeva (JP voice-actor) & 97.3\,\% & 2.7\,\% & 2.2 \\
ensemble SV-4 & 95.2\,\% & 4.6\,\% & 2.4 \\
ensemble all-6 & 95.0\,\% & 4.9\,\% & 2.2 \\
\bottomrule
\end{tabular}
\end{table*}

\begin{figure*}[!t]
\centering
\includegraphics[width=.85\textwidth]{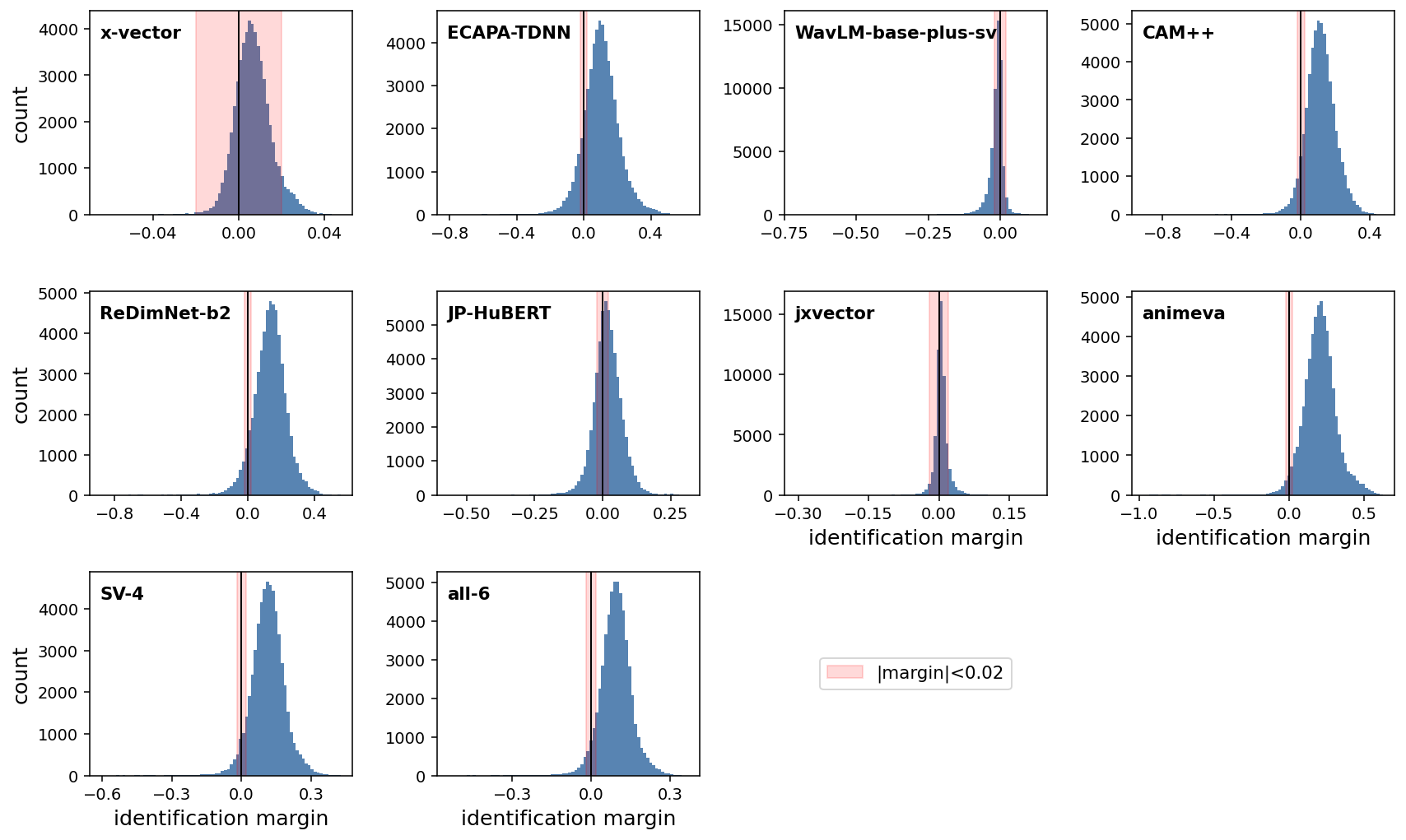}
\caption{Identification-margin distributions across the eight encoders and two ensembles: the signed margin \(S_{\mathrm{same}}\) \(-\) \(S_{\mathrm{diff}}\) per query; the mass near zero is the near-tie band of Section~\ref{subsec:raw_geometry}. Each panel is scaled to its own margin range, so the fixed $|$margin$| < 0.02$ band appears wider in the compressed-margin panels (x-vector, jxvector, WavLM).}
\label{fig:margins}
\end{figure*}

The confound and control-population comparisons (Section~\ref{subsec:domain_is_genuinly_high_density}) focus on representative encoders spanning the weak/strong and generic/domain-matched extremes (generic ECAPA-TDNN, domain-matched animeva, SV-4 ensemble). Even at high rank-1 accuracy, many queries are decided by a thin margin. Define the \emph{critical margin} as the fraction of queries for which $\left| S_{\mathrm{same}}-S_{\mathrm{diff}} \right| < 0.02$, where $S_{\mathrm{same}}$ is the highest-cosine score to the correct same speaker and $S_{\mathrm{diff}}$ that to the most confusable different speaker. The 0.02 is a diagnostic cutoff on the raw-cosine margin scale; for the strong verification encoders, the conclusion is stable across cutoffs 0.005--0.05 (protocol in Appendix~A). The absolute cutoff is \emph{not} comparable across encoders: x-vector, jxvector, and WavLM have order-of-magnitude-compressed margin distributions (std $\approx$0.009--0.03 vs $\approx$0.06--0.12 for the strong encoders), so the fixed band covers most of their mass and the near-tie fraction is meaningful only for the strong encoders. This near-tie band captures \textbf{\textasciitilde5.4--8.6\,\%} of queries for the strong generic single encoders (ECAPA-TDNN, CAM++, ReDimNet; animeva $\approx$1.9\,\%; full margin distributions in Fig.~\ref{fig:margins}, each panel scaled to its own margin range). Across all encoders, \textbf{hubness} --- the high-dimensional bias in which a few speakers become the nearest neighbor of disproportionately many utterances --- is consistently positive ($k$-occurrence skewness 1.4--3.6). The nominal permutation test gives $p \approx 0.005$ at the $B = 200$ floor of 1/201, so we report it as \textbf{exploratory/descriptive}: $p \approx 0.006$ under Benjamini--Hochberg \cite{benjamini1995controlling} but only $p \approx 0.07$ under the stricter Holm correction \cite{holm1979simple} across our 16-test family (10 per-encoder hubness tests + 3 confusion-graph structure tests $\times$ 2 encoders), i.e. it does not survive Holm. WavLM-base-plus-sv scores poorly at raw cosine (rank-1 27\,\%), most likely from representation anisotropy and/or our pooling of its output (it partly recovers under centering), so we treat it as a lower bound, not the model's true performance. We therefore treat \textbf{WavLM-base-plus-sv and the JP-HuBERT layer-mean as diagnostic probes, not tuned speaker-verification systems} (the latter a self-supervised representation without a verification head): they span the representation space, but the floor and hubness claims of Sections~\ref{subsec:misidentification_floor_survives} and~\ref{subsec:domain_is_genuinly_high_density} rest on the properly-configured verification encoders (ECAPA-TDNN, CAM++, ReDimNet, animeva) and their ensembles: a hubness or floor figure on a 27\,\%-rank-1 embedding is not, on its own, evidence about well-trained encoders.

\subsection{The misidentification floor survives calibration, ensembling, and discriminative re-ranking}
\label{subsec:misidentification_floor_survives}

The floor is not a bad threshold; it is geometry. To prove that, we separate what a threshold can move from what it cannot, reporting two identification-scale quantities. The first is the \textbf{closed-set rank-1 misidentification rate} (misID): the fraction of queries whose single nearest neighbor across the gallery belongs to a \emph{different} speaker (equivalently, margin $S_{\mathrm{same}} - S_{\mathrm{diff}} < 0$). Because every query's true speaker is enrolled and there is no reject option, this is a closed-set top-1 error (misID = 1 \(-\) rank-1), used deliberately as the ranking-quality summary that \emph{no monotone recalibration can change}. The second is the \textbf{$\bm{1:N}$ EER}: treating ``Is the query's top-scoring speaker the true speaker?'' as a threshold detector --- which \emph{does} carry a reject option --- we sweep the operating score and report the EER. This reject-capable, open-set-flavored metric is distinct from the pairwise speaker-verification EER. The evaluation is restricted to the codec-homogeneous subset of agency audio (official agency sources, all MP3, freelance/YouTube-derived excluded; this guarantees codec homogeneity only --- not same-studio or same-microphone conditions; Sections~\ref{subsec:domain_is_genuinly_high_density} and~\ref{sec:limitations_and_future_work}), with speaker-level bootstrap 95\,\% confidence intervals (CIs). We report all 10 encoders (Table~\ref{tb:asnorm}). The two diagnostic probes (WavLM, JP-HuBERT) are excluded from the floor claim (Section~\ref{subsec:raw_geometry}).

\begin{table*}[!t]
\centering
\caption{Raw $\rightarrow$ AS-norm effect per encoder (rank-1\,\%, $1:N$ EER\,\%, closed-set misID\,\%, $k$-occurrence hubness skew), on the codec-homogeneous agency subset. AS-norm roughly halves the \emph{threshold-placement} error ($1:N$ EER) and collapses hubness, but the \emph{ranking} error (misID) falls only modestly --- the residual floor. (The $1:N$ EER here is a \emph{mated}, max-over-gallery top-1-competitor characteristic, not comparable to a pairwise verification EER; see Appendix~A. The raw$\rightarrow$AS-norm comparison is within-encoder and unaffected. Bootstrap CIs in Appendix Table~\ref{tb:ci}.)}
\label{tb:asnorm}
\begin{tabular}{lrrrr}
\toprule
Encoder & Rank-1 (raw$\rightarrow$AS) & $1{:}N$ EER (raw$\rightarrow$AS) & misID (raw$\rightarrow$AS) & Hubness (raw$\rightarrow$AS) \\
\midrule
x-vector & 77.5 $\rightarrow$ 80.4 & 32.0 $\rightarrow$ 23.6 & 22.5 $\rightarrow$ 19.6 & 3.62 $\rightarrow$ 0.70 \\
ECAPA-TDNN & 85.7 $\rightarrow$ 87.4 & 27.0 $\rightarrow$ 15.9 & 14.3 $\rightarrow$ 12.6 & 2.18 $\rightarrow$ 0.80 \\
CAM++ & 91.8 $\rightarrow$ 93.3 & 18.7 $\rightarrow$ 10.3 & 8.2 $\rightarrow$ 6.7 & 1.82 $\rightarrow$ 0.78 \\
ReDimNet-b2 & 92.2 $\rightarrow$ 93.2 & 17.4 $\rightarrow$ 9.8 & 7.8 $\rightarrow$ 6.8 & 1.45 $\rightarrow$ 0.83 \\
jxvector & 65.9 $\rightarrow$ 69.8 & 44.2 $\rightarrow$ 32.8 & 34.1 $\rightarrow$ 30.2 & 1.45 $\rightarrow$ 0.43 \\
animeva & 97.5 $\rightarrow$ 97.8 & 10.1 $\rightarrow$ \textbf{3.9} & 2.5 $\rightarrow$ \textbf{2.2} & 0.90 $\rightarrow$ 0.86 \\
ensemble SV-4 & 94.8 $\rightarrow$ 95.6 & 15.6 $\rightarrow$ \textbf{7.0} & 5.1 $\rightarrow$ \textbf{4.4} & 1.57 $\rightarrow$ 1.07 \\
ensemble all-6 & 94.4 $\rightarrow$ 95.1 & 17.3 $\rightarrow$ 7.6 & 5.5 $\rightarrow$ 4.9 & 1.61 $\rightarrow$ 1.02 \\
\emph{WavLM (diagnostic)} & \emph{23.5 $\rightarrow$ 24.2} & \emph{60.4 $\rightarrow$ 71.0} & \emph{76.5 $\rightarrow$ 75.8} & \emph{1.41 $\rightarrow$ 0.34} \\
\emph{JP-HuBERT (diagnostic)} & \emph{61.0 $\rightarrow$ 64.3} & \emph{43.4 $\rightarrow$ 37.0} & \emph{39.0 $\rightarrow$ 35.7} & \emph{1.91 $\rightarrow$ 0.45} \\
\bottomrule
\end{tabular}
\end{table*}

\textbf{Takeaway:} AS-norm roughly halves the $1:N$ EER and collapses hubness, but closed-set misID falls only modestly (ensemble SV-4 5.1\,\% $\rightarrow$ 4.4\,\%). This asymmetry is exactly what the metric identities predict: AS-norm's test-side term is a monotone per-query shift that \emph{cannot} reorder neighbors, and its enrollment-side term reorders them only weakly (rank-1 improves $+0.3$ to $+3.9$\,pp, most for the weaker encoders, e.g., jxvector $+3.9$, x-vector $+2.9$), so calibration-invariance here is structural, not a finding. The load-bearing test is whether the floor survives \textbf{discriminative re-ranking that genuinely reorders neighbors}: LDA, WCCN, and a two-covariance PLDA log-likelihood ratio~\cite{brummer2010speaker}, train and test speakers are disjoint. Two back-end details keep the test fair (Appendix~A): LDA is run as a genuine dimensionality reducer, since a full-rank LDA coincides with WCCN under cosine, and the ensembles are reduced by principal component analysis (PCA)~\cite{jolliffe2002pca} before PLDA. We report pairwise-trial EER and the minimum and actual calibrated log-likelihood-ratio costs ($\mathrm{min}C_\mathrm{llr}$, $\mathrm{act}C_\mathrm{llr}$; their gap is the calibration loss, Appendix~A). All columns are averaged over three speaker-disjoint splits ($\pm$std) with a shared split-generation scheme; the misID column is a deterministic full-gallery quantity per split, so its cosine $\rightarrow$ LDA $\rightarrow$ WCCN $\rightarrow$ PLDA progression within a row is directly interpretable, while the pairwise EER/$\mathrm{min}C_\mathrm{llr}$/$\mathrm{act}C_\mathrm{llr}$ columns use independently seeded sampled trials and are comparable in trend, not trial-for-trial. \textbf{All Table~\ref{tb:reranking} values are over the $\bm{N = 497}$ evaluation-half speakers}, not the \textasciitilde1,100-speaker full gallery of Table~\ref{tb:asnorm}. This gallery-size difference itself lowers misID, so the AS-norm-4.4\,\%-to-PLDA-2.6\,\% ensemble drop mixes a back-end effect with an $N$ effect; the \emph{fixed-N} back-end effect is cosine 3.9\,\% $\rightarrow$ PLDA 2.6\,\% within Table~\ref{tb:reranking}.

\begin{table*}[!t]
\centering
\caption{Discriminative re-ranking of the misidentification floor (cosine~$\rightarrow$~LDA~$\rightarrow$~WCCN~$\rightarrow$~moment-fit two-covariance PLDA LLR), corrected back-ends, three speaker-disjoint splits. The final column gives the actual calibrated cost $\mathrm{act}C_\mathrm{llr}$ at the PLDA stage --- its closeness to $\mathrm{min}C_\mathrm{llr}$ is the small in-domain calibration loss.}
\label{tb:reranking}
\begin{tabular}{lrrrr}
\toprule
Encoder & $1{:}N$ misID\,\% (cos$\rightarrow$LDA$\rightarrow$WCCN$\rightarrow$PLDA) & Pairwise EER\,\% (cos$\rightarrow$PLDA) & $\mathrm{min}C_\mathrm{llr}$ (cos$\rightarrow$PLDA) & $\mathrm{act}C_\mathrm{llr}$ (PLDA) \\
\midrule
x-vector & 20.2 $\rightarrow$ 13.1 $\rightarrow$ 11.3 $\rightarrow$ \textbf{10.2} & 23.1 $\rightarrow$ 12.7 & 0.67 $\rightarrow$ 0.39 & 0.40 \\
ECAPA-TDNN & 11.3 $\rightarrow$ 9.6 $\rightarrow$ 9.6 $\rightarrow$ \textbf{9.0} & 15.7 $\rightarrow$ 10.9 & 0.50 $\rightarrow$ 0.35 & 0.35 \\
CAM++ & 6.1 $\rightarrow$ 4.4 $\rightarrow$ 4.5 $\rightarrow$ \textbf{3.8} & 13.4 $\rightarrow$ 8.6 & 0.45 $\rightarrow$ 0.29 & 0.29 \\
ReDimNet-b2 & 6.0 $\rightarrow$ 4.6 $\rightarrow$ 5.3 $\rightarrow$ \textbf{4.2} & 12.1 $\rightarrow$ 8.1 & 0.41 $\rightarrow$ 0.27 & 0.28 \\
jxvector & 27.8 $\rightarrow$ 13.9 $\rightarrow$ 12.5 $\rightarrow$ \textbf{15.9} & 26.9 $\rightarrow$ 16.6 & 0.76 $\rightarrow$ 0.52 & 0.52 \\
animeva & 1.6 $\rightarrow$ 1.5 $\rightarrow$ 1.4 $\rightarrow$ \textbf{1.4} & 7.7 $\rightarrow$ 6.6 & 0.28 $\rightarrow$ 0.23 & 0.23 \\
ensemble SV-4 & 3.9 $\rightarrow$ 3.0 $\rightarrow$ 2.9 $\rightarrow$ \textbf{2.6} & 12.4 $\rightarrow$ 7.1 & 0.40 $\rightarrow$ 0.24 & 0.24 \\
ensemble all-6 & 4.4 $\rightarrow$ 3.1 $\rightarrow$ 2.8 $\rightarrow$ \textbf{2.5} & 14.8 $\rightarrow$ 6.9 & 0.46 $\rightarrow$ 0.23 & 0.24 \\
\bottomrule
\end{tabular}
\end{table*}

Per-split std on the PLDA misID is $\leq$0.5\,pp for every encoder, and speaker-level bootstrap 95\,\% CIs (1{,}000 resamples over eval speakers) are tight and mutually well-separated: ensemble SV-4 2.6\,\% [2.45, 2.82], animeva 1.4\,\% [1.30, 1.55], ECAPA-TDNN 9.0\,\% [8.53, 9.50]. jxvector is a revealing anomaly: its pairwise EER \emph{improves} under PLDA (26.9\,\%$\rightarrow$16.6\,\%, that 16.6\,\% below LDA's and WCCN's operating points too), yet its argmax misID \emph{degrades} (12.5\,\% WCCN $\rightarrow$ 15.9\,\% PLDA). The effect is argmax-specific, not a simple generative mismatch. Consider the pairwise LLR $C + g_i + g_j + u_i^{\mathsf{T}}(q \odot u_j)$, where $u_i, u_j$ are the query and candidate embeddings in the simultaneously diagonalized PLDA space, $q$ is a fixed per-dimension weight vector, $\odot$ is the elementwise product, $C$ is a constant, and $g_i, g_j$ are quadratic per-vector self-terms (closed forms in Appendix~A). The query offset $g_i$ is constant across candidates and cannot change a $1:N$ argmax, but the \textbf{per-candidate self-term $g_j$} (a candidate-specific norm penalty) re-weights competitors and can flip the nearest-neighbor decision when many impostors sit in a narrow band --- precisely the high-density regime --- even as it improves calibrated pairwise separation. It does not bear on the floor claim, which rests on the properly-matched encoders and ensembles.

\textbf{Takeaway:} discriminative re-ranking lowers EER markedly and \textbf{does reduce} the misID floor (ECAPA-TDNN 11.3\,\% $\rightarrow$ 9.0\,\%, ensemble SV-4 3.9\,\% $\rightarrow$ 2.6\,\%, \textasciitilde30\,\% relative), but does \textbf{not} remove it: even under PLDA the residual stays several-fold above the control populations through the same matched-protocol back-end suite (matched $N$ and segments per speaker; PLDA-stage voice-actor (VA)/control misID ratios $\approx$ 1.9--6.1$\times$, diverging against neutral-read JVS --- Section~\ref{subsec:domain_is_genuinly_high_density}), and a floor persists across cosine, AS-norm, LDA, WCCN, and PLDA. For the domain-matched encoder, the back-ends are essentially flat (animeva 1.6\,\% $\rightarrow$ 1.5\,\% $\rightarrow$ 1.4\,\% $\rightarrow$ 1.4\,\%), consistent with a geometric rather than scoring-limited residual. So the floor is \textbf{re-ranking-resistant and geometry-limited}, not merely calibration-resistant: part is a scoring limit that re-ranking removes, and part is a property of the embedding geometry that no back-end in the standard linear/Gaussian suite removes. Nor is the residual an artifact of the moment estimator: an EM refit of the two-covariance model on the same splits and paired trials moves misID by at most 0.4\,pp on every encoder, and jxvector's argmax anomaly attenuates (Appendix~A).

\textbf{This also holds under discriminative \emph{nonlinear} re-rankers}: under the same protocol, two nonlinear scorers --- a neural PLDA~\cite{ramoji2020nplda} initialized at the two-covariance PLDA solution and fine-tuned with binary cross-entropy (BCE), and a pair-feature multilayer perceptron (MLP) over the PLDA-projected space --- neither lowered the misID floor below the linear PLDA (3 encoders, 3-split mean). For the neural PLDA, fine-tuning lowered the BCE loss, but held-out (train-speaker) validation misID rose monotonically, so model selection returned the initialization (= the moment-fit PLDA). The pair-MLP \emph{improves} pair EER/$\mathrm{min}C_\mathrm{llr}$ (SV-4 EER 7.1\,\% $\rightarrow$ 6.3\,\%, $\mathrm{min}C_\mathrm{llr}$ 0.24 $\rightarrow$ 0.22) yet \emph{worsens} the $1:N$ argmax misID --- the same pair-vs-argmax dissociation flagged for jxvector above. The only remaining untested lever is encoder fine-tuning, and the animeva-vs-generic gap shows representation choice can move the floor.

The same conclusion holds under \textbf{session-disjoint genuine trials}: running the identical suite with $S_{\mathrm{same}}$ restricted to segments from a different source recording (Table~\ref{tb:robustness} protocol, $N = 497$ splits), PLDA-stage misID stays at 13.0\,\% (SV-4), 13.9\,\% (animeva), and 25.8\,\% (ECAPA-TDNN) --- re-ranking closes only a small fraction of the same-session-to-cross-session gap, so the floor survives re-ranking in the harder, deployment-faithful condition too, and the domain-matched encoder's re-ranked advantage over the ensemble vanishes there.

\textbf{The floor persists under a standard centroid $\bm{1:N}$ protocol} (Table~\ref{tb:openset}). To confirm the floor is not an artifact of the segment-pool scorer, we also evaluate a conventional one-centroid-per-speaker gallery (\textasciitilde495 eval-half speakers, plain cosine, held-out probes; Appendix~A), reporting closed-set top-1 error and the open-set detection-and-identification rate (DIR) at a 1\,\%/5\,\% false-alarm rate (FAR). Absolute errors are higher than the PLDA segment-pool floor --- expected for an untrained cosine scorer over 8-segment centroids at $1{:}{\approx}495$ --- but both the floor and the encoder ordering (animeva $<$ SV-4 $<$ ECAPA-TDNN) are unchanged, so the geometry-limited floor is not a segment-pool artifact.

\begin{table}[!t]
\centering
\caption{Standard centroid $1{:}N$ companion metric (one L2-normalized 8-segment centroid per speaker, $N \approx 495$ eval-half gallery, plain cosine; mean over three splits). Closed-set top-1 error and open-set detection-and-identification rate (DIR) at 1\,\%/5\,\% false-alarm. The encoder ordering matches the segment-pool floor of Table~\ref{tb:reranking}; absolute values are higher because this is an untrained cosine scorer over $1{:}{\approx}495$ (Appendix~A).}
\label{tb:openset}
\footnotesize
\setlength{\tabcolsep}{4pt}
\begin{tabular}{lrrr}
\toprule
Encoder & \makecell{Closed-set\\top-1 err.\,\%} & \makecell{DIR@\\FAR=1\,\%} & \makecell{DIR@\\FAR=5\,\%} \\
\midrule
animeva & \textbf{25.1} & \textbf{47.4} & \textbf{59.6} \\
ensemble SV-4 & 40.4 & 28.3 & 39.2 \\
ECAPA-TDNN & 53.2 & 11.7 & 19.3 \\
\bottomrule
\end{tabular}
\end{table}

\begin{figure*}[!t]
\centering
\includegraphics[width=.60\textwidth]{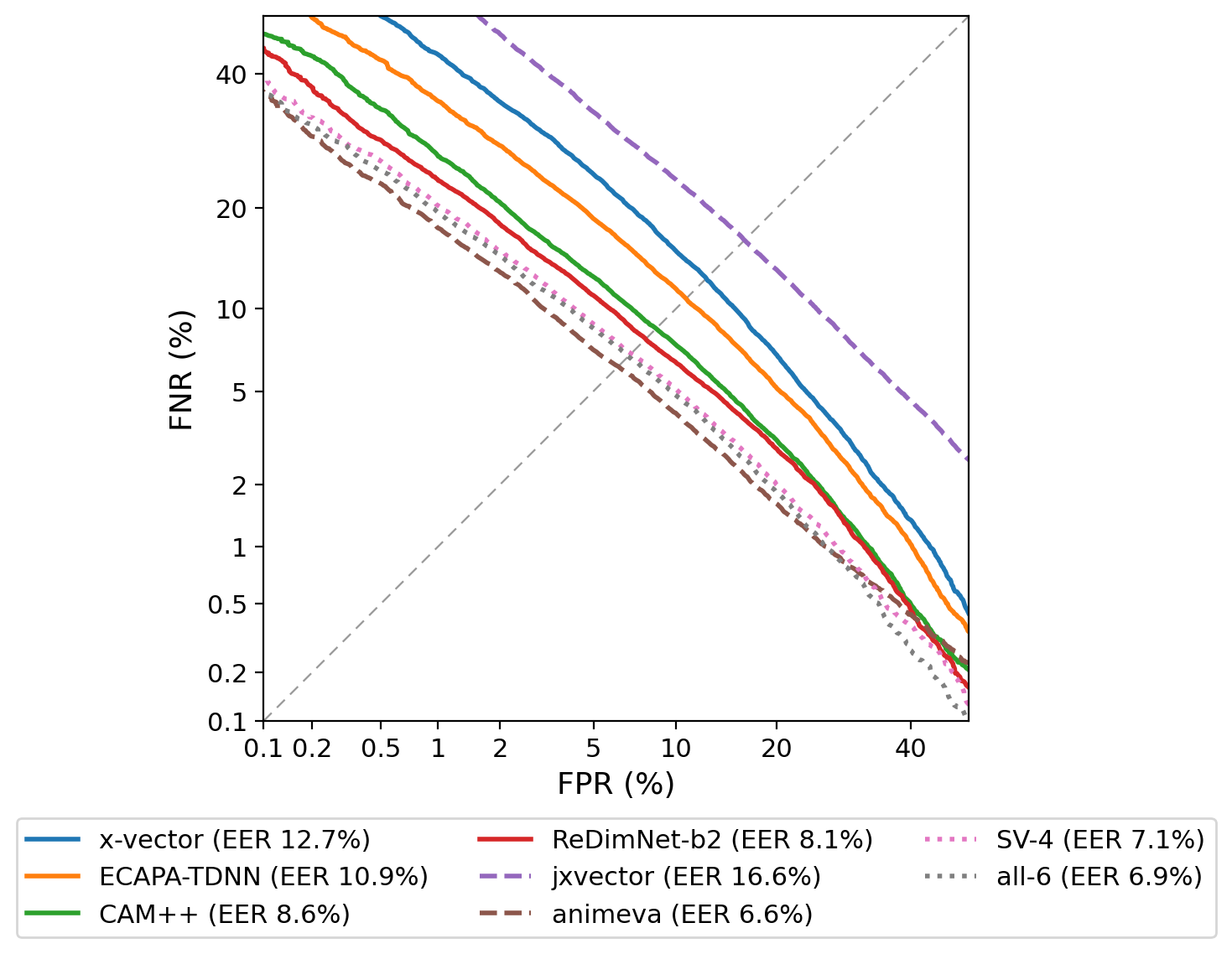}
\caption{Detection-error-tradeoff (DET) curves --- false positive rate (FPR, the false-alarm rate) versus false negative rate (FNR, the miss rate) --- on normal-deviate (probit) axes~\cite{martin1997det}, under the two-covariance PLDA-LLR (moment-fit) back-end, one curve per encoder (agency-only evaluation), from a single representative speaker-disjoint split; the scalar metrics of Table~\ref{tb:reranking} are three-split averages. The curves separate cleanly by encoder strength while every encoder retains a non-trivial operating region, consistent with the residual misidentification floor of Table~\ref{tb:reranking}.}
\label{fig:det}
\end{figure*}

On calibration, we separate two senses: \textbf{(i) the floor is a \emph{discrimination} limit, not a mis-set threshold} --- this rests on the \textbf{large, calibration-independent $\mathrm{min}C_\mathrm{llr}$} (0.23--0.52; $\mathrm{min}C_\mathrm{llr}$ is the best $C_\mathrm{llr}$ after \emph{optimal} calibration, so a high value cannot be a calibration artifact). The near-zero $\mathrm{act}C_\mathrm{llr}$ $-$ $\mathrm{min}C_\mathrm{llr}$ gap is \emph{consistent} but \textbf{not an independent calibration-adequacy test}, since $\mathrm{act}C_\mathrm{llr}$ is fit and scored on the same with-replacement, non-speaker-disjoint trial set, so a small gap is near-guaranteed (both are in-sample optimistic bounds --- Appendix~A). But \textbf{(ii) this calibration does not transfer across the synthetic-vs-real shift} of Section~\ref{subsec:defensive_clone_probe_and_clone_geometry}, where a real-calibrated threshold applied to clone audio fails --- a calibration-\emph{transfer} failure, distinct (DET curves: Fig.~\ref{fig:det}).

\textbf{What ``geometry-limited'' does and does not claim.} One could object that the floor is a \emph{label-structure} artifact rather than a property of the geometry: because each \texttt{speaker\_id} (our per-actor identity label) spans a voice actor's full performed range (narration, dialogue, character voices), some genuine trials pair a speaker's natural register with a very different character voice, so the residual might just be ``one performer is effectively several voices.'' First, this is the \textbf{operationally correct} definition, not a confound to remove: protecting a voice actor means protecting \emph{all} their performed voices, so a floor over the full intra-speaker range is exactly what a deployed attribution system faces. Second, restricting genuine trials to natural-voice-to-natural-voice pairs would likely lower the floor --- but would measure a different, narrower task: ``protect only the actor's natural voice.'' This paper measures protecting all performed voices, so the residual is a geometry-limited floor intrinsic to that task, not an artifact that vanishes under a different labeling choice (Section~\ref{sec:limitations_and_future_work}).

\subsection{The domain is genuinely high-density (controls + confounds)}
\label{subsec:domain_is_genuinly_high_density}

Is the domain hard, or does it just look hard due to a confound? It is genuinely hard. We bound the confounds by restriction and comparison with matched control populations.

\emph{Confounds bounded by restriction.} Even restricted to the codec-homogeneous subset of agency audio (Section~\ref{subsec:misidentification_floor_survives}; studio recording is plausible, but identical conditions are not guaranteed), the critical-margin and misID rates do not fall --- they slightly rise --- so they are not a codec artifact. Speakers from the same agency cluster only weakly on the main verification encoders --- a query's nearest neighbor is from the same recording source $\approx$ 1.6--1.7$\times$ chance for ECAPA-TDNN/ReDimNet, higher for other encoders (animeva 2.8$\times$, JP-HuBERT 3.0$\times$), so the residual channel effect is encoder-dependent --- and restricting impostors to \textbf{different agencies} leaves the critical-margin and misID rates essentially unchanged (ECAPA-TDNN critical-margin fraction $\approx$ 8.6\,\%). The thin margins are therefore a residual of genuine cross-studio voice proximity.

\emph{Comparison to control populations (matched on speakers and utterances per speaker, AS-norm, averaged over 3 seeds).} Voice actors are far harder to separate than the controls. In $1:N$ EER, they are $\approx$ 23$\times$ harder than JVS, a read-speech corpus of professional speakers (including voice actors and narrators) in a neutral reading condition; the contrast axis is style width, not profession (we use the \texttt{parallel100} subset of shared sentences). They are $\approx$ 2.6$\times$ harder than the same JVS with style variation added (JVS-varied: per speaker, 8 utterances from the non-parallel \texttt{nonpara30}, 2 from falsetto \texttt{falset10}, 2 from whispered \texttt{whisper10}), and 1.7--4.3$\times$ harder than Common Voice Japanese (CV; the general public). Adding style variation to the control thus collapses the ratio from $\approx$ 23$\times$ to $\approx$ 2.6$\times$ --- \textasciitilde91\,\% of the \emph{ratio} gap, though in absolute EER points inter-speaker proximity remains a substantial share. Intra-speaker style range is therefore the largest single driver of the difficulty --- \emph{intrinsic to the domain}, not a confound to remove --- yet voice actors remain harder than the general public. For the same reason, ``geometry-limited'' should \textbf{not} be read as ``distinct speakers are geometrically close'': the inter-speaker-proximity residual is a partial, not-fully-separated component (Contribution 3, Section~\ref{sec:limitations_and_future_work}). One encoder is excluded from these ratios: jxvector operates near ceiling on both populations (VA 45.7\,\% vs JVS-varied 42.1\,\% $1:N$ EER, ratio 1.1$\times$); at $\approx$ 46\,\% EER both saturate, and the ratio is uninformative.

The comparison is channel fair to the extent we can verify. The voice-actor side is \textbf{agency audio only} (official sources of speakers whose registry agency field is not ``freelance'', per Section~\ref{sec:data_and_pipeline}; freelance/YouTube-derived excluded) and is verifiably clean: a noise-floor signal-to-noise-ratio (SNR) proxy of 38--69\,dB, at or above the JVS studio reference (48\,dB) and well above crowd-sourced Common Voice (24\,dB). VA-vs-JVS is therefore a studio-vs-studio comparison, while the VA-vs-CV gap additionally includes a studio-vs-crowd-mic difference. That CV confound cuts both ways. Each CV speaker records on their own device, so genuine trials are channel-matched per speaker --- a device fingerprint that \emph{helps} the control and inflates the VA/CV ratio; conversely, crowd-mic degradation inflates the control's EER and makes the ratio conservative. We therefore treat the VA-vs-CV ratio as bounded by these opposing biases, and rest the ``genuinely harder'' claim on the JVS-varied comparison (a single shared studio, no per-speaker device fingerprint). Two caveats temper even this. First, the residual VA \textgreater{} JVS-varied gap is \textbf{not cleanly attributable to vocal-tract proximity}: professional demo reels share a broadcast mastering (loudness/EQ/limiting) aesthetic the SNR proxy misses, which could pull clean voices together independent of who is speaking (Section~\ref{sec:limitations_and_future_work}). Second, JVS-varied's phonation-mode proxy (nonpara/falsetto/whisper) is \textbf{not matched in kind} to the VA narration/dialogue/character range, if anything \emph{understating} the VA style range.

We also measured this comparison through the same back-end stages as Table~\ref{tb:reranking}, not only at the AS-norm stage (cosine $\rightarrow$ LDA $\rightarrow$ WCCN $\rightarrow$ PLDA; matched $N$ and segments per speaker; three speaker-disjoint splits, shared seeds). \textbf{At the PLDA stage the VA misID stays above 1$\times$ against every control on every encoder}: against neutral-read JVS the control side reaches 0\,\% misID (the ratio diverges; reported as 0 with the eval-probe counts); against Common Voice JA (312 speakers $\times$ 4 segments), $\approx$ 2--6.1$\times$; and even against style-varied JVS, $\approx$ 1.9--5.4$\times$. Speaker-level bootstrap confirms these gaps are not sampling noise: the VA and matched-control misID 95\,\% CIs are disjoint (e.g., ECAPA-TDNN VA vs Common Voice, a $\approx$6$\times$ misID ratio with non-overlapping intervals). Two caveats. Some cosine/WCCN-stage cells against style-varied JVS drop to 0.7--0.9$\times$ (control $\approx$ VA --- consistent with style width driving the difficulty), while the discriminative PLDA stage puts VA back above 1$\times$ in every cell. And the controls have few speakers (style-varied JVS has 34 train-side speakers, LDA/PCA dimensions capped at 33), so small-sample back-end instability affects some cells; the CV condition at 8 segments per speaker has only 11 train speakers and is unstable, hence the 312-speaker $\times$ 4-segment condition we use.

\subsection{Defensive clone probe and clone geometry}
\label{subsec:defensive_clone_probe_and_clone_geometry}

Now, the two harms the floor produces in practice: it points at the innocent, and it lets the guilty through. For clone generation, we use Seed-VC and the zero-shot TTS system Irodori-TTS at \textbf{120 targets each}, and the zero-shot TTS system GPT-SoVITS (v1--v4 and the v2ProPlus refinement) at 40 targets. Seed-VC conversions use six fixed source utterances from a single JVS speaker (jvs001; 3--4\,s read sentences), converted into each target via a concatenated (up to 12\,s) reference of the target's real segments; the same six sources are reused for the enrolled and non-mated targets (Appendix~A). Residual source-speaker traces in converted audio~\cite{cai2023sourcespeaker} are therefore a shared nuisance across all Seed-VC probes; two observations bound their impact: Irodori-TTS --- a zero-shot TTS with no source utterance at all --- shows the same qualitative harms, and the non-mated probe's real-audio baseline reproduces the false-alarm rates with no conversion involved. A direct enrichment control confirms this. Ranking the gallery centroids by similarity to the pooled jvs001 source vector, the wrongly-attributed actors are no closer to the source than chance on animeva (median rank 392/1,100; Mann-Whitney $p = 0.09$; none of the over-threshold wrongful accusations involves a jvs001-top-50 actor), while ECAPA-TDNN shows a small tail effect (12\,\% of nearest-neighbor (NN)-mismatches and 6/129 over-threshold events involve a jvs001-top-50 actor). Source leakage, therefore, cannot explain the wrongful-attribution phenomenon, contributing at most a minor fraction to the generic encoder, and nothing detectable on the domain-matched one.

One disclosure about GPT-SoVITS: it takes a transcript of the reference audio (prompt text) as input, and its no-reference-text mode degrades quality. Our pipeline auto-generated these transcripts using automatic speech recognition (ASR; faster-whisper small~\cite{radford2023whisper,fasterwhisper}, Japanese) without verifying accuracy (falling back on failure). ASR errors can only depress GPT-SoVITS's fidelity, so its tabulated detection rates provide a lower bound on what an attacker with accurate transcripts could achieve; we accordingly do \textbf{not} read GPT-SoVITS's miss rates as evidence of high-fidelity attacks --- Seed-VC and Irodori-TTS carry that claim.

We embed each clone and score it against a gallery of \textbf{\textasciitilde1,100 real-actor centroids} (every agency actor with $\geq$8 real segments) at a threshold \textbf{set at the real-vs-real EER operating point} --- the same condition as deployment. Enrolling each speaker by a single real-audio centroid collapses the intra-speaker style range (Section~\ref{subsec:domain_is_genuinly_high_density}), but a multi-vector enrollment (one vector per style) was also tested and leaves the conclusion unchanged (on animeva, Irodori-TTS attribution recall 67.5\,\% / NN-mismatch 22.4\,\%, vs 67.5\,\% / 22.6\,\% for the centroid method; $K = 8$ vectors per speaker, max-cosine scoring with the threshold re-derived at the multi-vector real-vs-real EER point, full 120-target Irodori-TTS set, $n = 720$).

\textbf{Scope of ``detection'' here.} The probe deliberately uses a bare speaker-embedding threshold --- the naive attribution rule this paper characterizes --- \textbf{not} a dedicated anti-spoofing countermeasure (CM). So \emph{detection} is \textbf{attribution recall} --- the clone's similarity to its \emph{true} actor is at or above the real-calibrated threshold (``Does the system still recognize the target?''), \textbf{not} synthetic-vs-real spoof detection; a ``miss'' is an attribution miss, not a CM false-negative. A modern self-supervised CM (e.g., AASIST/wav2vec2-based~\cite{jung2022aasist,tak2022wav2vec2cm}) is the right first stage of a deployed system (Section~\ref{sec:discussion}). We do not evaluate one, and its accuracy on this channel (Japanese demo-reel audio) and on these synthesizers --- none of which appear in any ASVspoof partition --- is unknown. Countermeasures degrade substantially in such a domain and under attack shift~\cite{muller2022generalize}, so deployment would likely require evaluation and possibly adaptation on matched synthesizers and channels.

Table~\ref{tb:clone_probe} reports attribution recall and two attribution-error quantities, in the order animeva \textbar{} ECAPA-TDNN, with speaker-level bootstrap confidence intervals in Appendix Table~\ref{tb:ci}. We distinguish two notions of false attribution:

\begin{itemize}

\item
  \textbf{NN-mismatch (nearest-neighbor mismatch, unconditional)}: the nearest real actor differs from the true actor. This counts \emph{every} clone, including sub-threshold ones a deployed detector would reject as ``no match'' --- so it overstates the operational harm of low-fidelity clones.
\item
  \textbf{Wrongful accusation (threshold-conditioned)}: the nearest real actor differs from the true actor \emph{and} is at or above the operating threshold --- an innocent actor both top-ranked and over the bar, the system acts on. This is the deployment-relevant event.
\end{itemize}

\begin{table*}[!t]
\centering
\caption{Defensive clone probe: attribution recall and false-attribution rates (animeva~$\vert$~ECAPA-TDNN). Bootstrap CIs in Appendix Table~\ref{tb:ci}.}
\label{tb:clone_probe}
\scriptsize
\setlength{\tabcolsep}{5pt}
\begin{tabular}{lrrrr}
\toprule
Method & \makecell{Attr.\\recall\,\%} & \makecell{NN-\\mismatch\,\%} & \makecell{\textbf{Wrongful-}\\\textbf{accusation\,\%}} & \makecell{Collateral\\(innocent $\geq$ thr)} \\
\midrule
Seed-VC (zero-shot VC, 2024, $n = 120$) & 81.8 \textbar{} 68.2 & 8.3 \textbar{} 28.6 & \textbf{1.5 \textbar{} 19.0} & 0.20 \textbar{} 3.70 \\
Irodori-TTS ($n = 120$) & 67.5 \textbar{} 51.0 & 22.6 \textbar{} 42.6 & \textbf{4.7 \textbar{} 17.6} & 0.24 \textbar{} 2.68 \\
GPT-SoVITS v1 $\rightarrow$ v4 & 2.9 $\rightarrow$ 64.2 \textbar{} 7.1 $\rightarrow$ 51.2 & 85.8 $\rightarrow$ 25.4 \textbar{} 82.9 $\rightarrow$ 40.0 & \textbf{2.9--10.0 \textbar{} 12.5--15.8} & 0.03 $\rightarrow$ 0.20 \textbar{} 0.69 $\rightarrow$ 0.72 \\
GPT-SoVITS v2ProPlus (v2 line) & 55.8 \textbar{} 46.2 & 30.4 \textbar{} 46.2 & \textbf{5.0 \textbar{} 17.5} & 0.27 \textbar{} 1.23 \\
\bottomrule
\end{tabular}
\end{table*}

\textbf{Takeaway:} the operational picture depends far more on the encoder than on the clone. On the off-the-shelf English encoder (ECAPA-TDNN), 12--19\,\% of clones actively name an innocent actor over the operating threshold --- across every clone type, including the low-fidelity GPT-SoVITS v1/v2 (their wrongful-accusation rate is non-monotone, hence a min--max range, not an arrow) --- while each Seed-VC clone places on average 3.7 innocent actors over the threshold. On the domain-matched encoder (animeva), wrongful accusations fall to 1.5--10\,\% and collateral to well under 1 actor per clone. The unconditional NN-mismatch column overstates low-fidelity harm: v1/v2 clones ``mis-attribute'' 74--87\,\% of the time only because they resemble \emph{no one} strongly, so their nearest neighbor is an arbitrary sub-threshold label a detector would report as ``no match,'' not an accusation. And the tabulated 12--19\,\% is a \textbf{lower bound}. The table's operating point is same-session-optimistic (genuine trials may share a source file with enrollment); the deployment-faithful \textbf{cross-file} point moves every quantity \emph{worse} --- detection rises \textasciitilde1--8\,pp, wrongful accusation \textasciitilde1--5\,pp (e.g., Seed-VC ECAPA-TDNN wrongful 19.0\,\% $\rightarrow$ 20.1\,\%). Both counts also grow with gallery size $N$, so a roster larger than our \textasciitilde1,100 would push them higher; and the non-enrolled-clone scenario (Section~\ref{sec:introduction}) is measured separately by the non-mated probe below.

\textbf{Clone fidelity and architecture.} The clones are competent, not degenerate, attacks. GPT-SoVITS spans two architecture lines (Section~\ref{sec:data_and_pipeline}), so \textbf{detectability tracks fidelity, and high fidelity is not exclusive to the newer architecture}: detection climbs v1$\rightarrow$v4 on animeva (2.9$\rightarrow$64.2\,\%), yet the v2-line v2ProPlus already reaches 55.8\,\% (46.2\,\% on ECAPA-TDNN), comparable to the v3/v4 line (59.2--64.2\,\% \textbar{} 44.2--51.2\,\%) --- the version arrow indexes fidelity, not a strict architectural ladder (on ECAPA-TDNN the two lowest-fidelity versions tie near the floor, v1 7.1\,\% vs v2 6.2\,\%). We use Seed-VC, not the lower-fidelity kNN-VC baseline \cite{baas2023knnvc}, because a higher-quality conversion is a \emph{stronger} attack and is more often detected. Our fidelity evidence is speaker similarity --- cosine to the target centroid, corroborated by an independent Resemblyzer encoder \cite{resemblyzer,wan2018ge2e} (Section~\ref{sec:discussion}, point (iii)) --- cross-checked by faster-whisper character error rate and UTMOSv2~\cite{baba2024utmosv2} DNN-predicted mean opinion score (DNN-MOS; Table~\ref{tb:robustness}), which shows the clones intelligible and not perceptually degenerate. UTMOSv2 is trained on English MOS corpora, out-of-domain for mastered Japanese demo-reel audio, so we read DNN-MOS only as ``clones are not degenerate,'' not a real-vs-clone naturalness comparison.

\textbf{Clone-reference / enrollment overlap (disclosure).} The reference audio each synthesizer clones from is drawn from the front of the target's segment list, overlapping the eight segments that build that target's enrollment centroid. Shared content, channel, and session can therefore inflate clone-to-target similarity, so the attribution-recall (``detect'') and wrongful-accusation figures for \emph{enrolled} targets are measured under a clone that partially reproduces the enrollment audio. This overlap is \emph{conservative} for the paper's central negative claim --- even with it, 18--32\,\% of high-fidelity clones fall below the real-calibrated threshold --- but \emph{optimistic} for absolute detection rates. Two existing controls bound parts of the effect --- the content-matched control (below) excludes the reference audio from the real side, and the cross-file operating point removes the same-source-file genuine trials from $\tau$ --- but a fully reference-disjoint clone probe remains open (Section~\ref{sec:limitations_and_future_work}). The non-mated probe is unaffected (its targets are not enrolled). Every harm rate here is a point estimate at one (EER or cross-file-EER) operating point; lowering the false-alarm rate (FAR) to suppress wrongful accusation only trades it for misses (Table~\ref{tb:tau}). No threshold satisfies both scenarios of Section~\ref{sec:introduction} --- the load-bearing claim (Section~\ref{sec:discussion}). Sweeping $\tau$ from the EER point to a 1\,\% real-vs-real FAR (Table~\ref{tb:tau}) confirms the trade-off is inescapable on the generic encoder: cutting ECAPA-TDNN wrongful accusation from 19.0\,\% to 0.3\,\% collapses its attribution recall from 68.2\,\% to 14.2\,\%, while the domain-matched animeva still recalls 57.1\,\% at near-zero wrongful accusation (Seed-VC clones). At every matched-FAR operating point and at EER, for both synthesizers, animeva simultaneously achieves higher recall \emph{and} equal-or-lower wrongful accusation than ECAPA-TDNN --- the operating curves never cross, so the domain-matched encoder dominates throughout, not merely at one point.

\begin{table*}[!t]
\centering
\caption{Attribution recall (Rec.) vs.\ wrongful accusation (Wr.) across operating points $\tau$ (real-vs-real FAR $\approx$ 1/2/5/10\,\% and the EER point of Table~\ref{tb:clone_probe}), in \%. At every point, animeva dominates ECAPA-TDNN on both axes.}
\label{tb:tau}
\begin{tabular}{lcccccccc}
\toprule
\multirow{4}{*}{Op.\ point} & \multicolumn{4}{c}{Seed-VC} & \multicolumn{4}{c}{Irodori-TTS} \\
\cmidrule(lr){2-5} \cmidrule(lr){6-9}
 & \multicolumn{2}{c}{animeva} & \multicolumn{2}{c}{ECAPA-TDNN} & \multicolumn{2}{c}{animeva} & \multicolumn{2}{c}{ECAPA-TDNN} \\
\cmidrule(lr){2-3} \cmidrule(lr){4-5} \cmidrule(lr){6-7} \cmidrule(lr){8-9}
 & Rec. & Wr. & Rec. & Wr. & Rec. & Wr. & Rec. & Wr. \\
\midrule
FAR $\approx$ 1\,\%  & 57.1 & 0.00 & 14.2 & 0.28 & 34.7 & 0.42 &  6.1 & 0.56 \\
FAR $\approx$ 2\,\%  & 62.5 & 0.14 & 18.8 & 0.69 & 39.3 & 0.42 &  8.9 & 0.97 \\
FAR $\approx$ 5\,\%  & 69.7 & 0.28 & 26.8 & 2.08 & 48.3 & 0.97 & 14.4 & 1.94 \\
FAR $\approx$ 10\,\% & 76.1 & 0.42 & 35.0 & 5.42 & 57.9 & 2.08 & 19.4 & 4.31 \\
EER                & 81.8 & 1.53 & 68.2 & 19.03 & 67.5 & 4.72 & 51.0 & 17.64 \\
\bottomrule
\end{tabular}
\end{table*}

When a clone \emph{does} land on the wrong actor, it does so \textbf{systematically}, not at random. The largest fraction of a target's misattributed clones concentrating on the same wrong actor (their \emph{consistency}) is 0.39--0.65 across synthesizers, versus a permutation baseline of $\approx$ 0.20--0.32 --- the correct null for a max-fraction statistic. That is a \textasciitilde1.3--2.5$\times$ enrichment (empirical $p < 0.001$ against both a uniform null over the full 1,100-actor gallery and a hub-preserving marginal null permuting the observed wrong-actor labels across targets). The pairwise version --- the probability that two clones of the same target hit the \emph{same} wrong actor --- is 0.06--0.40 versus a $\approx$1/1,100 uniform-chance rate. Comparing the max-fraction statistic to the 1/1,103 pairwise baseline would be the wrong null and overstate the enrichment; the effect is real but \textasciitilde2$\times$ chance, not \textasciitilde500$\times$ chance.

\textbf{Non-mated probe: false alarms from clones of non-enrolled speakers (direct measurement of scenario 1).} The 65 freelance actors in our corpus are not enrolled in the agency gallery (\textasciitilde1,100 centroids), so they serve as genuinely non-mated targets: we clone each with Irodori-TTS and Seed-VC (390 clones each) and score against the agency gallery at the same operating point. Since the target is not enrolled, every above-threshold hit is a false alarm --- a falsely accusatory misattribution of an innocent enrolled actor. Results (animeva~$\vert$~ECAPA-TDNN, speaker-level bootstrap 95\,\% CIs, at the real-audio EER operating point): 17.9\,\% {[}12.1, 24.4{]} \textbar{} 46.4\,\% {[}36.7, 56.4{]} of Irodori-TTS clones and 13.6\,\% {[}8.7, 19.2{]} \textbar{} 54.4\,\% {[}42.8, 64.4{]} of Seed-VC clones place their top-ranked enrolled actor above the threshold (at the cross-file point all figures worsen by \textasciitilde4--7\,pp; the number of innocent actors over the threshold is 0.17--0.25 per clone on animeva and 4.7--5.7 on ECAPA-TDNN, against a real-audio baseline of 3.9). \textbf{On the generic encoder, roughly half of the clones of people not enrolled at all falsely point to a real enrolled actor above the threshold}. The targets' own real audio (same freelance channel, $n = 756$) yields false alarms at nearly the same rates (17.5\,\% \textbar{} 48.9\,\%) --- so the false alarms stem from the gallery's density/geometry, not synthesis artifacts. Caveat: the freelance sources are a different channel from the agency gallery (YouTube/personal-site codecs), which can deflate similarities; this mainly affects the real-audio baseline (the clones carry the synthesizer's channel), so the real-audio rates may be optimistic (too low).

\textbf{Designed-voice probe: false matches with no cloning event.} Modern TTS can \emph{design} fictitious voices from a text description, with no reference audio and no target person \cite{guo2023prompttts}. We generated 60 designed voices with Irodori-TTS-600M-v3-VoiceDesign (caption-conditioned, no-reference mode) and scored them against the agency gallery in the same centered-cosine space and operating points as above. At the real-audio EER operating point, \textbf{11.7\,\% {[}95\,\% CI 5.0--20.0{]} (animeva) and 16.7\,\% {[}3.3--33.3{]} (ECAPA-TDNN) of the designed voices match an enrolled actor above threshold} (15.0\,\% {[}8.3--21.7{]} and 26.7\,\% {[}11.7--45.0{]} at the cross-file point; caption-level bootstrap over the 15 design prompts $\times$ 4 sampled candidates) --- false attributions with no victim-side cloning event. The score \emph{level} is diagnostic. On the domain-matched encoder, the flagged matches sit at the 31st percentile of the held-out genuine same-speaker distribution --- lower than a typical genuine match, consistent with ``coincidentally similar'', not ``a copy''. On the generic encoder, by contrast, they sit at the 65th percentile, \textbf{above the median genuine same-speaker pair}: by score alone, a designed voice flagged by ECAPA-TDNN is indistinguishable from a real match. No flagged match falls below the genuine 10th percentile on either encoder. One objection is that the generator's training data is undisclosed in the case of Irodori-TTS, so a flagged-designed voice could be a partial regurgitation of an enrolled actor rather than a coincidence. We cannot rule this out ($n = 60$, single generator, Section~\ref{sec:limitations_and_future_work}), but three observations argue against memorization as the dominant mechanism. (a) Genuinely non-enrolled \emph{humans} --- copies of no one --- false-alarm at comparable or higher rates in the non-mated probe (48.9\,\% on ECAPA-TDNN vs 16.7\,\% for designed voices), so gallery density alone suffices. (b) The matches are diffuse: no single enrolled actor absorbs more than a few flags, whereas regurgitation predicts repeated convergence on the same actors. (c) Most directly, the two encoders' flags are \textbf{disjoint} --- not one designed voice is flagged by both, and their top-1 neighbors agree on only 3.3\,\% of the 60 voices --- whereas a regurgitated actor would be recognized by both.

\textbf{Clone geometry (finding A).} Embedding all 2,640 clones (across the 7 synthesizer configurations of Table~\ref{tb:clone_probe}), they cluster by \textbf{target speaker} --- a clone's nearest neighbor is another clone of the same target \textasciitilde96--99\,\% of the time (ECAPA-TDNN 95.8\,\%, animeva 98.8\,\%), about 87$\times$ the \textasciitilde1.1\,\% chance rate for 120 targets, with silhouette coefficient~\cite{rousseeuw1987silhouettes} 0.21--0.26 --- but \textbf{not} by synthesizer (only \textasciitilde4.5$\times$ chance; silhouette $\approx$0). The embeddings clearly capture the speaker's identity of the clone. Yet \textbf{real-vs-synthetic audio separates linearly well above chance}: a speaker-disjoint GroupKFold~\cite{pedregosa2011sklearn} probe balanced across the seven synthesizers (the raw baseline of the channel control below, not the full 2,640) reaches balanced accuracy (chance = 50\,\%) of 86\,\% for ECAPA-TDNN and 73\,\% for animeva; a plain accuracy figure would be inflated by majority-class prediction. This indicates a systematic covariate shift: synthetic speech is distributed differently from real speech.

A \textbf{channel-control experiment} shows this shift is \emph{not explained by band-limiting or codec}: imposing the same 7 kHz band-limit and MP3 round-trip on both real audio and clones leaves separability essentially unchanged (ECAPA-TDNN 85.7\,\% $\rightarrow$ 86.5\,\%, animeva 72.9\,\% $\rightarrow$ 69.7\,\%; area under the receiver-operating-characteristic curve (AUC) 0.77--0.94 with balanced per-class recall). We stop short of calling the shift fully \emph{intrinsic}: this control removes only channel cues above 7\,kHz or destroyed by the codec, so sub-band microphone/room response could still contribute.

To test that, a \textbf{vocoder copy-synthesis control} analysis-resynthesizes the real audio through the \emph{identical} neural vocoder checkpoint Seed-VC v2 uses internally (BigVGAN v2~\cite{lee2023bigvgan}, 22\,kHz, 80-band, \texttt{nvidia/\allowbreak bigvgan\_\allowbreak v2\_\allowbreak 22khz\_\allowbreak 80band\_\allowbreak 256x}), measuring exactly the vocoder stage's possible traces. Resynthesized audio from 200 speakers (800 segments) preserves speaker identity almost perfectly (top-1 match to the speaker's own real centroid 98--100\,\%; centroid cosines nearly identical to real audio), and its linear separability from real audio is only \textbf{58--62\,\% balanced accuracy} --- far below the real-vs-clone separability (73--86\,\%). Moreover, a separator trained on real-vs-clone flags the copy-synthesized audio as ``synthetic'' only 16--21\,\% of the time, barely above the same speakers' real-audio control (13--18\,\%; both on the 174 speakers held out of the separator's training, so out-of-sample). A narrow vocoder-resynthesis trace thus exists but is small, and is not the main driver of the real-synthetic shift. (The separator is a linear probe in the \emph{speaker-embedding} space, not a CM; the point is not undetectability but that, in the geometry attribution relies on, whatever displaces clones off the manifold is not vocoder traces.)

By contrast, a \textbf{cross-synthesizer generalization test} (train on some synthesizers, test on a held-out one) \emph{does} degrade accuracy --- held-out Seed-VC/Irodori-TTS fall to 53--71\,\% versus 80--93\,\% in-distribution --- so part of the shift is synthesizer-specific, and our separability probe is an in-distribution upper bound for any embedding-level gate (a dedicated CM is a different, stronger detector in-distribution, but inherits the same generalization risk~\cite{muller2022generalize}). This resolves the apparent paradox in Section~\ref{subsec:defensive_clone_probe_and_clone_geometry}: clones are \emph{relatively} faithful (a clone-vs-clone target probe reaches 92.5\,\% on animeva) but their \emph{absolute} position is shifted off the real-speaker manifold, so a real-calibrated threshold either drops them below $\tau$ or lands them beside the wrong real centroid. \textbf{The shift's size is also heterogeneous across synthesizers:} low-fidelity GPT-SoVITS v1/v2 land far off-manifold (per-method separability 0.84--0.97), whereas high-fidelity Seed-VC sits closer to the manifold on the domain-matched encoder (separability 0.84--0.93, its clones embedding near the target's real centroid). So for the \emph{best} attacks, the detection failure is driven as much by the real-calibrated operating point as by a large shift, while for the worst attacks the shift dominates.

\begin{figure*}[!t]
\centering
\includegraphics[width=.85\textwidth]{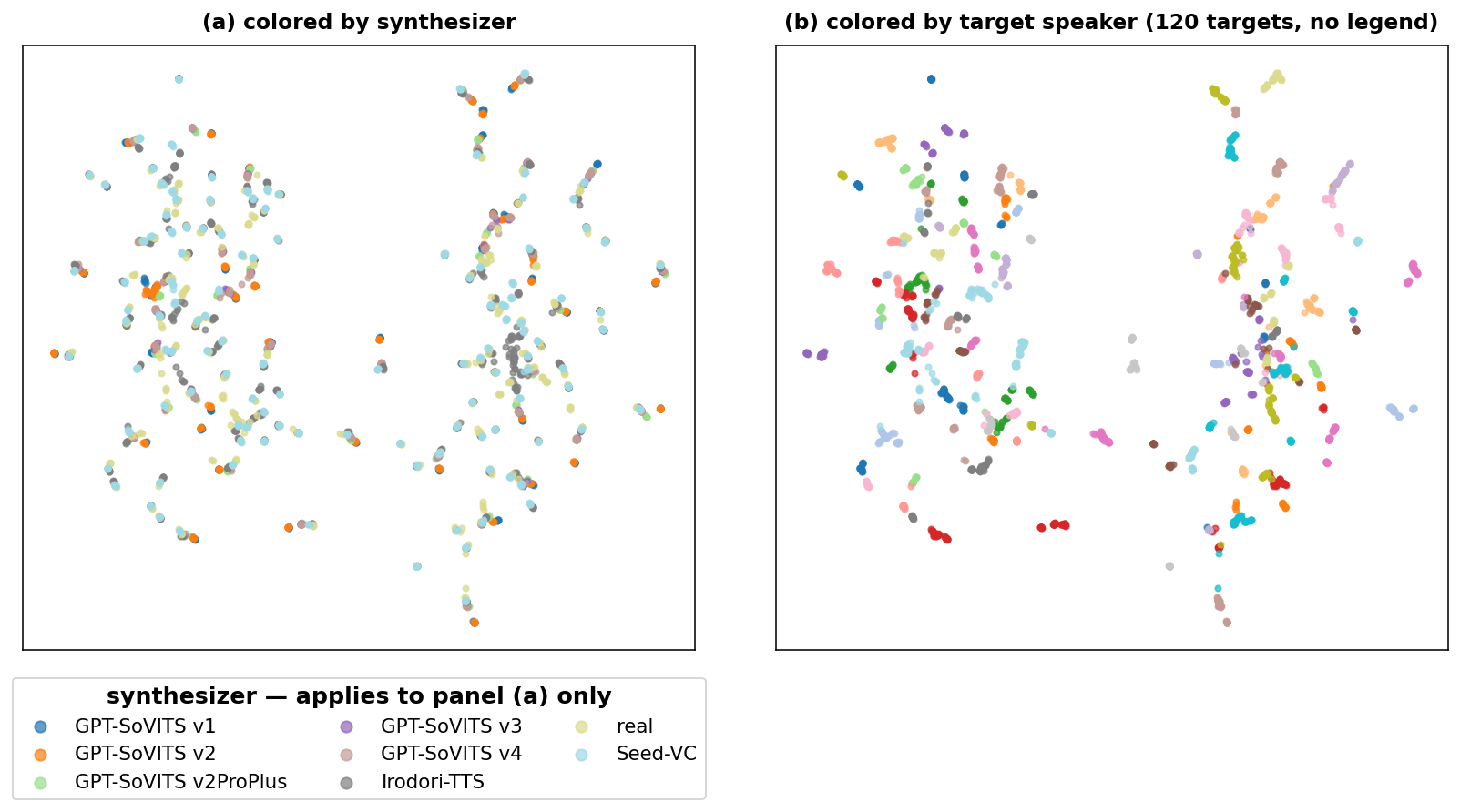}
\caption{Two-dimensional UMAP~\cite{mcinnes2018umap} projection (cosine metric, $n_{\mathrm{neighbors}}{=}15$, min\_dist $=0.1$, random\_state $=0$; centered and L2-normalized embeddings) of clone and real-speech embeddings: clones cluster by \emph{target speaker} (panel b) rather than by \emph{synthesizer} (panel a). Real-speech points are overlaid (the ``real'' class, panel a); the synthetic-vs-real offset of Section~\ref{subsec:defensive_clone_probe_and_clone_geometry} is not visible in this 2-D projection and is quantified separately by the linear-separability probe (73--86\,\% balanced accuracy).}
\label{fig:clone_geometry}
\end{figure*}

\textbf{In short, the detector's failure is consistent with a covariate shift (and threshold calibration), not missing identity information} (Fig.~\ref{fig:clone_geometry}). We say \emph{consistent with} rather than \emph{proven}, because the separator cannot isolate a single cause. The cross-synthesizer drop to 53--71\,\% shows a large \textbf{synthesizer-specific} component --- which, by the copy-synthesis control, is not explained by vocoder-resynthesis traces alone, and plausibly includes acoustic-model traces.

A \textbf{phonetic/content-distribution} difference, meanwhile, is bounded by a \textbf{content-matched real/clone pair control}: we generated Irodori-TTS clones speaking the transcribed text of 477 held-out real segments (120 targets; identical recipe and reference audio; Appendix~A). Against these content-matched clones, separability is nearly unchanged from the content-\emph{un}matched originals (ECAPA-TDNN 88.6\,\% vs 89.9\,\%, animeva 80.7\,\% vs 84.8\,\%), so content mismatch is at most a minor contributor. The same holds for the other synthesizers: for Seed-VC we use \textbf{self-conversion} --- converting the target's own held-out real segment to their own voice, matching content \emph{and} prosody --- and separability is nearly unchanged (ECAPA-TDNN 89.6\,\% vs original 93.5\,\%, animeva 81.9\,\% vs 87.0\,\%), so neither content nor prosody explains the shift; GPT-SoVITS v4 is likewise nearly unchanged (ECAPA-TDNN 81.4\,\% vs 84.2\,\%, animeva 79.2\,\% vs 75.7\,\%). After the copy-synthesis and content-matched controls across three synthesizers, the covariate-shift reading is best supported; the only remaining reservation is that the copy-synthesis control uses a single vocoder family (Section~\ref{sec:limitations_and_future_work}).

\subsection{Cross-cutting findings (worse on the generic encoder; animeva advantage verified on held-out speakers)}
\label{subsec:cross_cutting_findings}

\begin{itemize}
\item
  \textbf{Fairness (finding B).} Gender labels here are \textbf{agency-published presented-gender categories} (scraped from gendered roster pages), \textbf{not self-identified gender}, and are \textbf{binary only} --- at collection time, no agency or registry we surveyed published a non-binary category. They cover 458/1,168 speakers, a non-random, agency-selected subsample. On ECAPA-TDNN (English/VoxCeleb), female actors are misidentified \textbf{24.5\,\% (95\,\% CI {[}22.5, 26.7{]}) versus 6.3\,\% {[}5.4, 7.3{]}} for male actors (speaker-level bootstrap; 244 female / 214 male) --- a \textbf{$+18.2$\,pp gap, CI {[}16.0, 20.5{]}, one-sided $p < 0.001$}. Under animeva the gap is \textbf{absent}: female 2.1\,\% {[}1.7, 2.6{]} vs male 2.9\,\% {[}2.3, 3.5{]} (\(-\)0.7\,pp, CI {[}\(-\)1.5, \(-\)0.0{]}). We separate \textbf{existence} from \textbf{magnitude}. The \emph{existence} of an ECAPA-TDNN gap is robust: the two encoders score the \emph{identical} gallery, so a pure gallery-size effect (which would raise both) cannot explain a gap large on ECAPA-TDNN and null on animeva. It also \textbf{survives within every style stratum with enough labeled speakers} (ECAPA-TDNN female\(-\)male misID $+10.5$\,pp within narration and $+22.1$\,pp within dialogue, both $p < 0.001$; style-stratified on the same speakers --- name-read/full/freeform buckets dropped for $<5$ speakers per gender), so it is \textbf{not} a style-composition artifact. The \emph{magnitude} ($+18.2$\,pp) is, however, genuinely confounded: misID rises with the number of same-gender competitors a query faces, and the two-encoder contrast does not equalize the gallery's gender composition (710 speakers are gender-unlabeled). The point value is therefore gallery-dependent, not a portable effect size. To guard against reading the closure as mere floor-compression --- at animeva's \textasciitilde10$\times$ lower base error, an \emph{absolute} pp gap shrinks mechanically even if the disparity persisted --- we report the scale-free \textbf{female/male misID ratio}: 3.9$\times$ on ECAPA-TDNN (24.5/6.3) versus \textbf{0.74$\times$} on animeva (2.1/2.9). The disparity does not merely shrink, it \emph{reverses direction} --- which compression alone cannot produce. Still, animeva \emph{closing} the gap remains an \textbf{existence proof, not an isolated cause}: its in-domain training, its (unknown) training-set gender composition, and its architecture are confounded (Section~\ref{sec:limitations_and_future_work}). An overlap audit against the published VNDB voice-actor metadata for animeva's training set finds that 246/1,168 (21\,\%) of our speakers are seen. The effect is small (animeva misID 2.1\,\% seen vs 2.8\,\% unseen), and \textbf{the closure holds on the name-unmatched speakers}: over the 360 with a gender label (181 female / 179 male), animeva scores female 2.4\,\% / male 2.9\,\% (gap \(-\)0.5\,pp, 95\,\% CI {[}\(-\)1.3, +0.4{]} --- no female disadvantage, point estimate slightly reversed), versus ECAPA-TDNN female 23.3\,\% / male 6.0\,\%. The 21\,\% is a \textbf{lower bound from exact name matches} only: source-level leakage (shared agencies/studios) is not excluded, so ``name-unmatched'' is a weak negative. Less-famous speakers (at or below the median credited works) are also more confusable (17.1\,\% vs 12.9\,\% on ECAPA-TDNN). We analyze gender, fame, and agency marginally, not jointly, and the 39\,\% gender-label coverage is itself an agency-selected subsample (Section~\ref{sec:limitations_and_future_work}).
\item
  \textbf{Style-dependent misattribution risk (finding C).} On ECAPA-TDNN, \textbf{scripted-dialogue} segments are more confusable than narration (misID 18.3\,\% vs 13.1\,\%) --- the misattribution risk depends on speaking style. This concerns which audio styles the defender should judge and enroll, not impersonation-attack design. We label styles from demo-reel track names, which mark narration vs dialogue vs name-reads but do \textbf{not} isolate \emph{character voices} (these fall inside the dialogue and unlabeled ``other'' buckets), so we report a narration-vs-dialogue effect and make no separate character-voice claim (Section~\ref{sec:limitations_and_future_work}). By contrast, animeva, trained on in-domain acted speech, separates dialogue from narration well (1.8\,\% vs 4.4\,\%).
\item
  \textbf{Cost of a single operating point (finding D).} Using one common threshold for all speakers, ECAPA-TDNN leaves \textbf{63\,\% of speakers with a false-reject rate (FRR) above 20\,\% and 47\,\% with a false-accept rate (FAR) above 20\,\%}; per-speaker optimal thresholds spread across 0.60--0.75. Adding an \textbf{abstain option} (deferring the least-confident 50\,\% of queries by \textbar margin\textbar) cuts misID from 13.3\,\% to 4.4\,\% (Fig.~\ref{fig:threshold_spread}).
\item
  \textbf{Confusion graph (finding E).} We build a confusion graph in which each utterance has an edge to its ``most confusable different speaker'' (its nearest impostor segment), attributed to that segment's speaker. Two structural properties are robustly significant; one earlier claim does not survive a stricter null:

\begin{figure*}[!t]
\centering
\includegraphics[width=.94\textwidth]{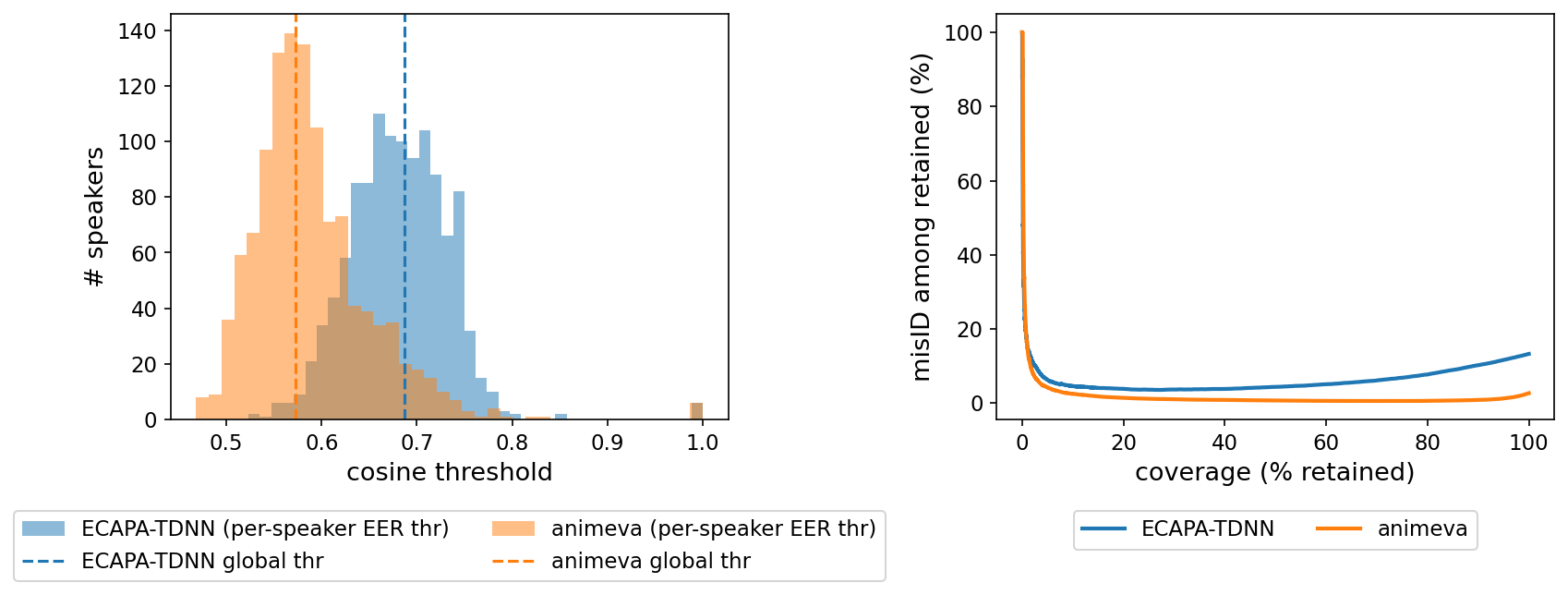}
\caption{The per-speaker optimal-threshold spread --- thresholds range widely (0.60--0.75 on ECAPA-TDNN), so one global threshold cannot serve all speakers --- and the misID risk--coverage curve for an abstain option that defers the least-confident queries by margin; deferring the least-confident 50\,\% cuts ECAPA-TDNN misID from 13.3\,\% to 4.4\,\%.}
\label{fig:threshold_spread}
\end{figure*}

  \begin{itemize}
  
  \item
    \textbf{Community structure.} Modularity~\cite{newman2004modularity} --- how cleanly the network divides into communities --- is
    \textbf{0.42 vs 0.12 in a degree-preserving rewired null} (ECAPA-TDNN; animeva 0.45 vs 0.14), a
    configuration-model null rewiring edges at random while preserving each node's degree (observed and null graphs are unweighted, so the double-edge-swap~\cite{maslov2002specificity} is a clean degree-preserving comparison); empirical permutation $p \approx 0.005$ (floored at 1/201 by the number of rewirings). These are voice-type clusters, not noise.
  \item
    \textbf{Reciprocity.} Confusions are far more \emph{mutual} than chance: only \textbf{56\,\% of edges are one-directional, vs 94\,\% under the null} (empirical $p \approx 0.001$). If \emph{A}'s nearest impostor is \emph{B}, then \emph{B}'s is disproportionately often \emph{A} --- genuine bidirectional confusability.
  \item
    \textbf{In-degree hubness --- retracted as a segment-count artifact.} The raw in-degree distribution is skewed (skewness 1.44; some speakers are the nearest impostor of many others), but edges point at nearest impostor \emph{segments}, so a speaker with more enrolled segments has proportionally more chances to be a target even with no voice-level hubness. Against a \textbf{size-proportional null} (targets drawn with probability proportional to segment count), the observed skew of 1.44 matches the null's 1.45 (empirical $p = 0.54$ on ECAPA-TDNN; on animeva, the observed skew is \emph{below} the null, $p = 1.0$). We therefore \textbf{do not claim in-degree hubness beyond what the segment-count imbalance produces}; the uniform-null $z$-score (17.5) was inflated by exactly this confound. The \emph{segment-level} hubness of Section~\ref{subsec:raw_geometry} ($k$-occurrence skew, a per-segment property not driven by per-speaker segment counts) is separate and unaffected.
  \end{itemize}

  So the confusion structure is real and significant in its \textbf{community and reciprocity} (who-confuses-whom clusters into mutually-confusable voice-type groups), but it is \textbf{not} a story of a few ``super-hub'' actors absorbing disproportionately many false matches once speaker size is controlled (Fig.~\ref{fig:confusion}).
\end{itemize}

\begin{figure*}[!t]
\centering
\includegraphics[width=.94\textwidth]{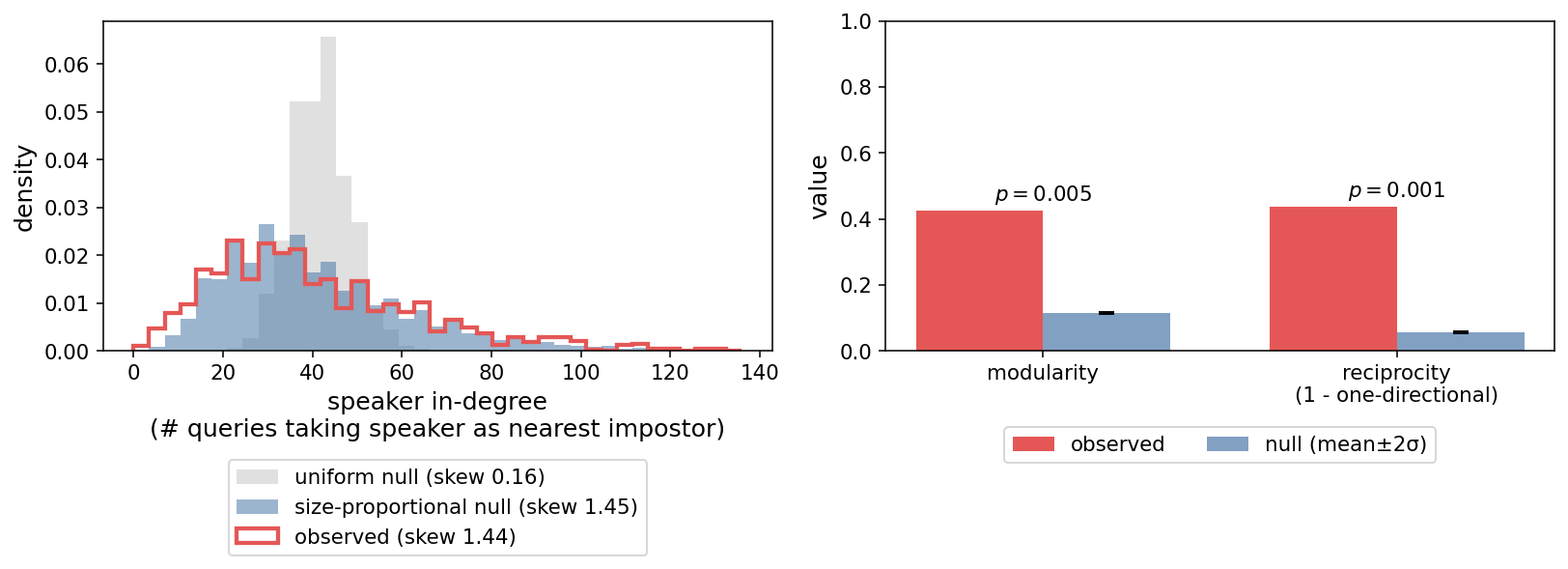}
\caption{Confusion-graph structure: the observed speaker in-degree distribution overlaps the size-proportional null (in-degree hubness is a segment-count artifact, and is retracted), while modularity and reciprocity remain far above their nulls (ECAPA-TDNN; the animeva comparisons are given in the text of finding E).}
\label{fig:confusion}
\end{figure*}

\subsection{Robustness: The floor and the shift are not confound artifacts}
\label{subsec:robustness}

A skeptic can ask, of every result above, whether it is an artifact of some confound rather than the embedding geometry. We ran the controls; none altered the conclusions. Table~\ref{tb:robustness} collects them, with detailed protocols in Appendix~A and the sections cited per row; each row states the confound, the test, and the outcome.

\begin{table*}[!t]
\centering
\caption{Confound controls for the misidentification floor (F) and the real-versus-synthetic shift (S).}
\label{tb:robustness}
\footnotesize
\setlength{\tabcolsep}{4pt}
\renewcommand{\arraystretch}{1.2}
\begin{tabular}{p{0.14\textwidth}p{0.26\textwidth}p{0.36\textwidth}p{0.15\textwidth}}
\toprule
Potential confound & Test & Result & Conclusion \\
\midrule
Codec differences (F, S) & Restrict to the codec-homogeneous agency MP3 subset (Section~\ref{subsec:domain_is_genuinly_high_density}) & Critical-margin / misID do not fall (slightly rise) & Unchanged \\
Recording channel / band (S) & Same 7\,kHz low-pass + MP3 round-trip on real and clone audio (Section~\ref{subsec:defensive_clone_probe_and_clone_geometry}) & Real-vs-clone separability essentially unchanged (ECAPA-TDNN 85.7\,\% $\rightarrow$ 86.5\,\%) & Unchanged \\
Vocoder fingerprint (S) & Copy-synthesize real audio through Seed-VC's BigVGAN family, re-embed (Section~\ref{subsec:defensive_clone_probe_and_clone_geometry}) & Real-vs-copysynth 58--62\,\% $\ll$ real-vs-clone 73--86\,\%; identity preserved (top-1 98--100\,\%) & Vocoder trace small; shift is more than the vocoder \\
Content / phonetic mismatch (S) & Content-matched real/clone pairs, three synthesizers; Seed-VC self-conversion matches content \emph{and} prosody (Section~\ref{subsec:defensive_clone_probe_and_clone_geometry}) & Separability nearly unchanged (Seed-VC 89.6 vs 93.5\,\%; Irodori-TTS 88.6 vs 89.9\,\%; GPT-SoVITS v4 81.4 vs 84.2\,\%) & Content/prosody do not explain the shift \\
Scoring back-end (F) & Run controls through the full cosine $\rightarrow$ LDA $\rightarrow$ WCCN $\rightarrow$ PLDA suite at matched $N$ (Section~\ref{subsec:domain_is_genuinly_high_density}) & PLDA-stage VA misID stays >1$\times$ every control (neutral JVS \(\infty\), CV 2--6.1$\times$, style-varied JVS 1.9--5.4$\times$) & Several-fold gap is measured, not inferred \\
Nonlinear scoring (F) & Neural PLDA and a pair-feature MLP on the PLDA space (Section~\ref{subsec:misidentification_floor_survives}) & Neither lowers misID below linear PLDA (3 encoders, 3 splits) & Floor survives nonlinear re-ranking \\
Gallery size $N$ (F) & misID-vs-$N$ curve, $N \in {100\ldots{}1103}$, 10 draws each (Section~\ref{sec:limitations_and_future_work}) & misID grows with $N$ but the encoder ordering and floor level hold (animeva 0.98\,\% at $N=100$; ECAPA-TDNN 6.97\,\%) & Floor is not an artifact of large $N$ \\
Background music (F, S) & Audit with an AudioSet tagger (PANNs CNN14): BGM in ~18.6\,\% of agency files; re-run excluding flagged files (Section~\ref{sec:limitations_and_future_work}) & misID floor persists (shifts $\leq$ 1.4 pt, down on the full gallery, mixed-sign on the codec-matched subset); separability does not fall (ECAPA-TDNN 85.7$\rightarrow$85.8\,\%, animeva 72.9$\rightarrow$78.6\,\%) & BGM was noise on the real class, not a driver \\
Duplicate / split identities (F) & Cross-speaker centroid audit; drop the higher-ID member of every near-duplicate pair (Section~\ref{sec:limitations_and_future_work}) & Raw-cosine misID floor essentially unchanged (animeva 2.54\,\% $\rightarrow$ 2.41\,\%) & Not manufactured by duplicates \\
Mastering loudness (F) & Re-embed after ITU-R BS.1770 loudness normalization (Section~\ref{sec:limitations_and_future_work}) & Floor, hubness, same-agency enrichment unchanged (misID $\leq$0.08 pt; SV-4 5.05\,\% $\rightarrow$ 5.07\,\%) & Loudness is not the confound (EQ/limiting undecidable, Section~\ref{sec:limitations_and_future_work}) \\
Same-session genuine trials (F) & Restrict every genuine trial to a different source file (912 speakers with $\geq$2 files; Section~\ref{sec:limitations_and_future_work}) & Session-disjoint misID rises to 25.2\,\% (SV-4), 18.6\,\% (animeva), 38.5\,\% (ECAPA-TDNN); AS-norm no longer rescues it, and the full re-ranking suite ($N{=}497$ splits) still leaves PLDA-stage misID at 13.0\,\% (SV-4), 13.9\,\% (animeva), 25.8\,\% (ECAPA-TDNN) & Same-session tables are a substantial lower bound; the floor survives re-ranking session-disjoint \\
Fusion method (F) & Learned stacking / per-encoder weights on the SV-4 members, speaker-disjoint (Section~\ref{sec:data_and_pipeline}) & Learned weights only marginally below mean-pool (3.5\,\% vs 3.8\,\%) via ECAPA-TDNN pruning; stacking worse (4.2--4.6\,\%) & Floor survives learned fusion \\
Clone naturalness/intelligibility (S) & UTMOSv2 DNN-MOS (5-fold ensemble) and faster-whisper CER on clones vs real audio (Section~\ref{sec:limitations_and_future_work}) & CER 1.4--4.3\,\%; DNN-MOS comparable to the real demo-reel audio (real 2.59; Seed-VC 2.96, Irodori-TTS 2.61, GPT-SoVITS 2.19--2.42) & Clones intelligible, not perceptually degenerate \\
\bottomrule
\end{tabular}
\end{table*}

The one control we cannot run decisively is mastering normalization: professional demo reels share a broadcast loudness, equalization (EQ), or limiting aesthetic, and without the pre-master audio, a linear normalization cannot cleanly separate that shared mastering transfer function from genuine vocal-tract spectral cues (Section~\ref{sec:limitations_and_future_work}). As a \emph{suggestive} check, we re-embedded the codec-matched subset after ITU-R BS.1770~\cite{itu1770} loudness normalization: the floor, hubness, and same-agency enrichment are essentially unchanged (misID moves $\leq$0.08 pt on every encoder --- e.g., SV-4 5.05\,\% $\rightarrow$ 5.07\,\%, ECAPA-TDNN 14.33$\rightarrow$14.33\,\% --- as expected, since the encoders are near gain-invariant). This rules out loudness but not the full mastering transfer function (EQ/limiting), which is not linearly invertible without the pre-master; that remains a genuine limitation, not a decisive control.

\section{Discussion}
\label{sec:discussion}

One thread runs through all of this: a \textbf{generic (English/VoxCeleb) embedding with a single operating point degrades systematically} in this high-density Japanese voice-actor domain --- across calibration, fairness, style, thresholding, and confusion structure alike --- and each degradation is materially reduced by a domain-matched (Japanese, voice-actor-trained) encoder. But mitigation is not elimination: even the best-calibrated, domain-matched ensemble leaves a residual, geometry-limited misidentification floor and hub structure.

For detecting unauthorized clones and training-data misuse, a single operating-point attribution cannot avoid both wrongful accusations and missed attributions. The \textbf{misidentification/missed-detection} harm is distributed unequally: it concentrates on female, less-famous, and densely-confusable-community speakers (Section~\ref{subsec:cross_cutting_findings}, findings B and E). These are marginal, non-interaction-modeled effects on the victim-side misID event; we did \textbf{not} measure per-gender wrongful accusation. We instead recommend an open-set attribution pipeline \textbf{in the spirit of SASV} (extending SASV's binary target/spoof decision to $1:N$ attribution): (1) a real anti-spoofing countermeasure as a gate --- a self-supervised learning (SSL)-based CM such as AASIST/wav2vec2, which we do \textbf{not} evaluate. Our 73--86\,\% separability probe (Section~\ref{subsec:defensive_clone_probe_and_clone_geometry}) is only an in-distribution, optimistic stand-in for such a gate, not a deployed CM: it drops to \textasciitilde53--71\,\% on held-out synthesizers, and to \textbf{$\approx$53.5\,\%, essentially chance}, for held-out Seed-VC on the generic encoder --- and SASV/ASVspoof \cite{jung2022sasv,yamagishi2021asvspoof} show CMs generalize poorly to unseen synthesizers. When the synthetic nature of the audio is uncontested, this gate is unnecessary and only stage (2) operates --- the gate matters precisely when synthetic-vs-real is contested. Then (2) attribution with a domain-matched encoder using \textbf{per-speaker calibration, synthetic-shift-corrected enrollment, and an abstain region} routed to human review. Because the SASV finding \cite{jung2022sasv} that an integrated system beats a naive cascade also applies here, we frame this cascade as a performance lower bound. Given that wrongful accusation reaches 12--19\,\% on a generic encoder and generalization to unseen synthesizers is weak, such a system suits \textbf{detection support (flagging audio for human verification), not autonomous enforcement} (takedowns or licensing penalties) --- the abstain-and-human-review region is essential, not optional. The abstain region defers exactly the coincidental-over-inclusion cases; whether a human resolves them is untested here, and the designed-voice scores (flagged matches at the 65th percentile of genuine same-speaker pairs on the generic encoder) suggest it is hard, as does the lay-listener evidence of Section~\ref{sec:related_work}. Neither the machine nor the unaided ear is adjudicative on its own --- trained forensic examination remains unmeasured here --- so the decision they feed must remain one piece of evidence among others. These are design requirements motivated by, but not validated in, this study: no end-to-end tandem metric is reported.

\textbf{Provenance standards are complementary but do not address false attribution.} Content-provenance standards such as C2PA \cite{c2pa2024} --- increasingly adopted by voice-synthesis vendors, sometimes with proprietary watermarking \cite{elevenlabs2026synthid,elevenlabs2024iconic,respeecher2024,microsoft2023personalvoice,openai2024voiceengine} --- cryptographically bind a signed manifest to authorized synthetic audio. This is a \emph{positive} proof of origin, but asymmetric: the \emph{absence} of a manifest proves nothing, and manifests are metadata that transcoding strips routinely (soft-binding to re-associate a stripped manifest is not yet robust for audio). Waveform watermarking (e.g., AudioSeal \cite{sanroman2024audioseal}) survives transcoding better but only marks compliant generators. Provenance therefore addresses neither failure mode here: it cannot stop an unsigned clone --- or a natural voice --- from being \emph{falsely attributed} to an enrolled actor (Section~\ref{subsec:defensive_clone_probe_and_clone_geometry}), and does not protect a victim whose unauthorized clone carries no manifest. Nor is the asymmetry only technical: a licensing contract with actor \emph{A} documents the clone's \emph{true source} but does not rebut a similarity-based claim that the output is ``\emph{B}'s voice'', and such documentation may be under non-disclosure agreement (NDA) or absent entirely (a self-clone, an informally consented recording, a designed voice). Provenance and similarity-based attribution are orthogonal: the former certifies the authorized channel, while the risks measured here live in the unsigned remainder.

\textbf{Broader implication.} Our measurements support one concrete implication: acoustic similarity --- whether computed by an encoder or a voiceprint procedure --- is an unreliable identity substrate in this domain, better suited to \emph{screening} than to a stand-alone decision criterion. Our results speak only to \emph{machine} similarity; the available human evidence (Section~\ref{sec:related_work}) suggests a human-identifiability standard would not inherit the machine's over-inclusion so much as substitute population-dependent failures of its own \cite{hayashi2026vtuberfans}; a matched-foil test remains open (Section~\ref{sec:limitations_and_future_work}). We develop the fuller policy argument in separate work.

\textbf{Why a single operating point cannot satisfy both requirements.} The two scenarios of Section~\ref{sec:introduction} are the two error types of one operating point over the geometry we characterize, so they trade off rather than resolve together. \emph{(i) False attribution is independent of consent}: wrongful accusation runs 12--19\,\% between enrolled targets and roughly half for clones of non-enrolled speakers, so even a consented, licensed clone of \emph{A} can land nearest a different real actor \emph{B}; and 12--17\,\% of \emph{designed} fictitious voices with no reference audio also match an enrolled actor, so false attribution need not involve a cloning event at all. \emph{(ii) The mis-attributed target is predictable}: the confusion graph is community-structured and reciprocal (Section~\ref{subsec:cross_cutting_findings}, finding E), so which actor \emph{B} absorbs a false match is structurally determined, with no ``super-hub'' actor once speaker size is controlled. \emph{(iii) A high-similarity clone can fall below threshold}: a clone can score high on an independent similarity measure (Seed-VC clones reach Resemblyzer 0.80--0.85) yet sit below a real-calibrated point, because of the synthetic-vs-real covariate shift. \emph{(iv) The error is unequally distributed}: it concentrates on female and less-famous speakers (finding B; measured victim-side). Loosening the threshold worsens (i)--(ii); tightening it worsens (iii) --- no single operating point escapes the trade-off, the technical reason an abstain-and-human-review design is preferable to autonomous enforcement.

\section{Ethics and reproducibility}
\label{sec:ethics_and_reproducibility}

Purpose is defensive (measuring detectability); \textbf{no audio is redistributed and no clone is deployed, offered as a service, or used to impersonate}. Article 30-4 (the information-analysis limitation)~\cite{japan_copyright_act}, which Art. 102(1) extends to performers' neighboring rights, covers our collection/analysis use, \emph{subject to its proviso}: the limitation does not apply where the use ``unreasonably prejudices the interests of the copyright owner.'' We assess that it does not here, on documented facts: no collected audio or clone is redistributed, deployed, offered as a service, or used to impersonate; the clones are internal measurement inputs only; no per-speaker attack recipes are published; and the released artifacts are derived statistics, not audio --- so the use neither substitutes for nor competes in the market for the actors' performances or their licensing. This assessment is consistent with the Agency for Cultural Affairs' March 2024 ``General Understanding on AI and Copyright''~\cite{aca2024aiguidance}. Whether website terms of service can contractually override Art. 30-4 is an open question acknowledged in that guidance; our provenance log records robots.txt compliance for the registry site, an identified User-Agent with a contact address, and rate-limited fetching. Collection, analysis, and storage were entirely within Japan, and the data is neither transferred to nor redistributed to other jurisdictions; we therefore treat Japanese law as the governing framework and do not attempt an exhaustive multi-jurisdictional analysis. Copyright limitations cover \textbf{neither} publicity/personality rights --- a separate, case-law-only interest in Japan (Section~\ref{sec:introduction}) --- nor personal-information law (below). We therefore \textbf{anonymize individuals in all per-speaker findings} (e.g., hub examples are reported as ``one speaker'', not by name) and report aggregate geometry, not per-speaker attack recipes. Provenance (URL, timestamp, SHA-256) is logged for every fetch. For reproducibility, the released repository pins \textbf{dependency versions} (torch, torchaudio, transformers, speechbrain) and tabulates the \textbf{source IDs and exact revisions} of every encoder and clone system --- the Seed-VC, GPT-SoVITS, and Irodori-TTS repository commits, the HuggingFace and ModelScope model revisions, and the torch.hub checkpoints (ReDimNet, jxvector) --- together with random seeds and speaker-disjoint split definitions. \textbf{To keep the release consistent with the anonymization pledge}, all released per-speaker artifacts --- derived statistics and split definitions --- are keyed by \textbf{salted one-way hashes of the registry ID, not the seigura ID or name}, with the ID $\leftrightarrow$ hash mapping withheld (available to reviewers/replicators on request under a data-use agreement administered by the author: requests via the contact address in the header; research use only, no re-identification, no redistribution, verification only). Critically, the clone-probe outputs name \emph{innocent third parties} (the wrongfully-attributed actor per clone), so we release these only as \textbf{aggregate rates}, not per-clone rows resolvable to a named actor. An animeva training-speaker hash set (already one-way) is released to enable the overlap audit. \textbf{Hash-keying anonymizes the identifier column, not the embeddings themselves}: speaker embeddings are biometric identifiers, and nearest-neighbor matching against public audio would in principle re-identify most entries --- our own rank-1 accuracies (up to 97.3\,\%, Table~\ref{tb:raw_geometry}) indicate such linkage would succeed for most speakers (an in-principle attack indicated by the rank-1 results, not a measured re-identification rate). We therefore treat the centroids as \emph{identity-linked, not anonymized}, and---consistent with this very finding---do \textbf{not} release them publicly; they are available to reviewers/replicators only on request under a data-use agreement, for verification; embedding inversion \cite{champion2021invertibility} is a further residual consideration. Any listed actor (or their agency) may request removal of their derived artifacts, and removals will be honored in all future releases.

\textbf{Personal-information law (APPI).} Japan's Act on the Protection of Personal Information (Act No.~57 of 2003, as amended)~\cite{appi2003} applies alongside copyright: voice-feature data extracted so as to identify a specific individual plausibly meets the designation of an \emph{individual identification code} (APPI Cabinet Order Art.~1(i)), so the per-speaker embeddings --- linked via a withheld mapping to named registry entries --- constitute personal information, and salted hashing of the key does not by itself remove them from APPI scope. As an independent researcher, the author does not clearly fall within the academic-research-institution exceptions and does not rely on them; the artifacts are instead handled under the general obligations --- purpose limitation to this research, security control, no public release of the embeddings, and no third-party provision except verification-only access for reviewers/replicators under the data-use agreement, and the removal mechanism above.

\textbf{No human-subjects research is part of this paper.} All results are computational analyses of existing public audio and of synthetic audio generated for this study; no experiment involving human participants was conducted (a human-perception listening study is mentioned only as future work in Section~\ref{sec:limitations_and_future_work}). As secondary analysis of already-public recordings plus researcher-generated synthetic audio --- with no intervention or interaction with living individuals and no private information obtained through intervention --- it does not meet the human-subjects definition that triggers institutional review-board oversight, and none was applicable. This mode of study is standard in speaker recognition, whose principal benchmarks (e.g., VoxCeleb) are themselves assembled from public recordings without individual consent; our handling (anonymization, withholding of the biometric embeddings, aggregate-only sensitive outputs, and a removal mechanism) is more restrictive than that norm.

\textbf{Clone generation and consent.} The voice clones analyzed here were generated \emph{solely} as internal measurement inputs --- from a bounded subset of the roster (\textasciitilde40--120 enrolled targets per probe, plus 65 non-enrolled speakers) --- to quantify the very attribution harms this paper documents; the actors did not consent to being cloned, and we do not treat the absence of objection as consent. We hold the resulting tension to a minimum: no clone is redistributed, deployed, offered as a service, or used to impersonate; clone outputs are reported only as aggregate rates, never as per-clone rows resolvable to a named actor; no per-speaker attack recipe is published; and any listed actor or their agency may request removal (honored in all future releases). We judge that generating a bounded set of non-deployed clones to measure --- and thereby help defend against --- unauthorized cloning is proportionate to that defensive purpose, but we state the tension explicitly rather than treat it as resolved.

\textbf{Ethics declaration.} As an unaffiliated researcher, the author had no institutional review board available; the study was nonetheless designed to meet the substantive standards such review protects --- data minimization, non-maleficence, no redistribution of audio or clones, aggregate-only reporting of sensitive outputs, and removal on request --- and to comply with the legal and data-protection analysis above (Japanese Copyright Act Art.~30-4 and its proviso, APPI general obligations, and publicity/personality-rights considerations). No experiment involving human participants was conducted (as stated above). The author welcomes ethics correspondence at the contact address in the header.

\textbf{Provenance of the domain-matched encoder, and how to obtain one cleanly.} The public animeva encoder we use as our domain-matched control is itself reported~\cite{animeva} to be trained on the VisualNovel\_Dataset ($\approx$6.95M clips derived from commercial visual novels, distributed under a hex-obfuscated name and without the voice actors' consent). We flag this as an ethical hazard rather than endorse it: we use animeva \textbf{only as a research artifact} to test \emph{whether} a domain-matched encoder can mitigate the failures we document, neither redistribute the dataset nor recommend deploying this specific model, and report every contamination-sensitive claim on both animeva and the contamination-free ECAPA-TDNN. Crucially, domain matching is the mitigating factor, not this particular training set: \textbf{for an operator who genuinely wants the mitigation, training (or licensing) a domain-matched encoder on consented voice-actor audio is advisable} --- a protection system resting on non-consensually collected data of the very rights-holders it is meant to protect undermines its own legitimacy, its acceptance by those parties, and its credibility as evidence --- and this is entirely feasible with studio cooperation. That no \emph{clean, consent-based} domain-matched encoder is currently available off the shelf is itself a finding, and an argument for building one. (animeva also has author-acknowledged implementation quirks --- batch normalization (BatchNorm)~\cite{ioffe2015batchnorm} $\rightarrow$ group normalization (GroupNorm)~\cite{wu2018groupnorm}, non-standard input scaling --- so we treat it as a useful but imperfect community artifact, not a gold standard.)

\textbf{Conflict-of-interest declaration.} The author is an independent researcher who also provides contracted engineering services to a company operating text-to-speech products. This research was conceived, funded, and executed independently, with no involvement in, access to, or review of the study by that company. The author declares no other competing interests relevant to this work.

\section{Limitations and future work}
\label{sec:limitations_and_future_work}

\begin{itemize}

\item
  \textbf{Human identification is out of scope.} We measure machine attribution only (Section~\ref{sec:ethics_and_reproducibility}). The one familiar-listener measurement \cite{hayashi2026vtuberfans} used one target, one trial per condition, fixed order, and foils far apart in embedding space (cosine $\approx$0.18); foils drawn from the near-neighbour communities of Section~\ref{subsec:cross_cutting_findings} are the natural next probe --- does the human floor track the geometric one --- left to future work.
\item
  \textbf{Clone-probe scale and cross-lingual scope:} the clone probe uses \textasciitilde40--120 targets, one registry, one language; cross-lingual/out-of-registry and cross-system replication are future work. Training-set independence is not guaranteed: the corpora of Irodori-TTS and GPT-SoVITS are undisclosed and unaudited against our roster, so contamination is possible --- asymmetrically, making the miss and wrongful-accusation figures conservative while rendering attribution-recall and fidelity figures potentially optimistic. Seed-VC --- identity via converting a fixed source through the target reference --- is the relatively safer fidelity anchor, though its corpus is likewise unaudited.
\item
  \textbf{Identification metric is closed-set; the clone probe is mated-probe.} Our headline misID/rank-1 is a \emph{closed-set} top-1 error, reported alongside a reject-capable $1:N$ EER; the clone probe adds a reject threshold but is still \textbf{mated} (not open-set; no open-set CMC~\cite{decann2013cmc}). The non-enrolled false-alarm scenario is instead \textbf{directly measured} by the non-mated probe across the 65 out-of-gallery freelance speakers, where, on the generic encoder, roughly half of the non-enrolled clones point to an enrolled actor above threshold. Re-scored as $1:1$ verification, the harm persists: a randomly named innocent actor is falsely matched at the base rate ($2.8$/$10.1\,\%$, animeva/ECAPA-TDNN), but the actor a clone most resembles at $8$--$43\,\%$ --- \emph{above} the $1:N$ rate --- even as victim recall rises to $93$--$99\,\%$. Falling below the threshold is likewise not proof of non-identity.
\item
  \textbf{Segment-pool gallery and gallery size.} The core misID scores each query segment against every other \emph{segment}, so \(S_{\mathrm{diff}}\) is a maximum over tens of thousands of impostor segments and grows with the per-speaker segment count and gallery size $N$; we control for this ($N$-matched comparisons), but absolute misID values are $N$-dependent, not comparable across tables at different $N$ (497 eval-half, \textasciitilde1,100 full, 100/455 controls). The $N$ dependence is quantified by the misID-vs-$N$ curve and duplicate audit in Table~\ref{tb:robustness} (floor and encoder ordering hold across $N$); the audit covers agency but not the \textasciitilde65 freelance IDs.
\item
  \textbf{No anti-spoofing countermeasure or adaptive attacker.} The clone probe uses a bare embedding threshold, not a deployed CM (Section~\ref{subsec:defensive_clone_probe_and_clone_geometry}), and every clone is an \emph{honest} clone of its own target: the 12--19\,\% wrongful-accusation rate is an \textbf{incidental} floor at a single global operating point, not the output of an adversary optimizing a clone toward a chosen victim or to evade a gate. An SSL-based CM + SASV tandem and an adaptive-attacker analysis (aimed at the defense's robustness limits, not attack recipes) are future work; the present numbers are an attacker-agnostic lower bound, with the wrongful-accusation-vs-$\tau$ trade-off reported in Section~\ref{subsec:defensive_clone_probe_and_clone_geometry}.
\item
  \textbf{Clone fidelity is speaker-similarity only.} We verify clone fidelity by cosine to the target centroid plus an independent Resemblyzer encoder, cross-checked with an automated intelligibility and naturalness axis (Table~\ref{tb:robustness}), confirming the clones are not perceptually degenerate; ``high-fidelity'' therefore denotes ``high speaker-similarity'' pending a human listening study. Seed-VC and GPT-SoVITS are 2024-era public systems, and Irodori-TTS is a 2026 release; the arsenal does not exhaustively cover the latest commercial zero-shot or flow-matching systems, so generalizability to them remains untested.
\item
  \textbf{Contamination of the Japanese control encoders is unaudited.} The overlap audit covers animeva only. jxvector is trained on YouTube-scale JTubeSpeech and JP-HuBERT on \textasciitilde19,000 h of Japanese TV-broadcast audio (ReazonSpeech~\cite{yin2023reazonspeech}); prominent voice actors may be present in both, so we cannot exclude that part of their behavior is memorization; the ``language-confound isolated by a Japanese but non-VA encoder'' reading is provisional until a JTubeSpeech overlap audit is run.
\item
  \textbf{Speaker identity is a performer label, not a verified single natural voice.} The corpus is built from agency demo reels showcasing an actor's \emph{range} (narration, dialogue, character voices), so a single \texttt{speaker\_id} can span very different performed registers that the enrollment centroid averages over. We treat this intra-speaker range as domain-intrinsic, not a confound, but do not verify that same-speaker genuine pairs are the same natural voice vs two character performances, nor isolate \emph{character voices} as a labeled style (they fall in the dialogue/``other'' buckets; Appendix~A). A within-natural-voice genuine-trial condition and a character-voice label are future work.
\item
  \textbf{Production/mastering homogeneity.} Restricting impostors to different agencies removes a shared studio \emph{chain} but not the industry-wide broadcast mastering aesthetic, which could pull clean voices together independent of vocal-tract proximity. A loudness-normalized re-run rules out loudness (Section~\ref{subsec:robustness}), but EQ/limiting is not linearly invertible without the pre-master, so this stays a genuine limitation.
\item
  \textbf{Fairness magnitude is not covariate-adjusted.} The $+18.2$\,pp gender gap is not measured on a gender-balanced impostor gallery, and gender, fame, and agency are analyzed marginally, not jointly; the gap's \emph{existence} and survival within style strata are robust (finding B), but its magnitude and a covariate-adjusted coefficient are future work.
\item
  \textbf{Same-session genuine trials.} Segments are consecutive windows from the same source files, so genuine trials are not cross-session; this inflates target scores, making absolute EER/misID an optimistic (lower) bound --- conservative for the ``domain is hard'' thesis, \emph{asymmetrically} so since JVS-varied's genuine pairs are deliberately cross-style. The session-disjoint re-run confirms this (Table~\ref{tb:robustness}): the clone probe is directly bounded at a cross-file operating point.
\item
  \textbf{Clone reference audio overlaps enrollment audio.} For enrolled targets, the synthesizers' reference segments overlap the enrollment centroid, inflating absolute detect rates (conservative for the ``clones are missed'' claim); this is bounded in Section~\ref{subsec:defensive_clone_probe_and_clone_geometry} (content-matched control, cross-file operating point), and a fully reference-disjoint probe is future work.
\item
  \textbf{animeva advantage is confounded with domain match.} The Japanese voice-actor encoder wins on every axis, but we do not isolate \emph{why}: in-domain training, female-VA representation, and architecture are entangled. The advantage holds on held-out speakers (so not pure memorization), but we cannot separate ``domain-matched training'' from ``fairer-by-design'' --- a controlled comparison and representation probing are future work. animeva is best read as a \textbf{mitigation existence proof}, not a finished recipe.
\item
  \textbf{Generalizability:} one registry, one language, 63.5\,\% coverage with public-sample selection bias of \textbf{uncertain direction} --- distinctive voices could make it \emph{under}-estimate difficulty, hub-prone newcomers \emph{over}-estimate it. Coverage is also \textbf{agency-concentrated} (\textasciitilde45 adapters, \textasciitilde18\,\% of 255 agencies), over-representing larger agencies. Cross-lingual/out-of-registry checks are future work.
\end{itemize}

\section{Conclusion}

An embedding space can look accurate and still fail. In this high-density professional-speaker domain, it retains a \textbf{geometry-limited} misidentification floor and segment-level hubness that the high headline accuracy masks. Calibration only shifts the operating point; a monotonic recalibration cannot reduce a rank-1 error at all. Ensembling, and re-ranking that genuinely reorders neighbors --- LDA, WCCN, two-covariance PLDA (moment-fit and EM), and nonlinear re-rankers (neural PLDA, a pair-feature MLP) --- both reduce the floor, but only partly --- to \textasciitilde2.6\,\% at best, and \textasciitilde13\,\% under deployment-faithful session-disjoint trials. The remainder is a property of the embedding geometry relative to these back-ends; encoder fine-tuning is the untested lever. Because these are the two error types of one operating point, no single threshold escapes both: loosening it accuses the innocent, tightening it misses real clones. In this domain, then, fixed-threshold attribution of voice clones or training data is unreliable --- and, on a generic encoder, unfair. The path forward is not a better threshold but a different design: spoofing-aware speaker verification extended to open-set attribution, with an anti-spoofing gate (not evaluated here), a domain-matched encoder, per-speaker calibration, and an abstain option. Such a system can support human review; on this evidence, it cannot enforce on its own.

\section*{Appendix A. Evaluation protocol and metric definitions}

\textbf{Embeddings and scoring space.} Each segment is L2-normalized. Per-speaker enrollment centroids are the mean of the speaker's first eight agency segments; for the score-normalized (AS-norm) and clone-probe analyses, centroids and query vectors are additionally mean-centered by the gallery mean and re-normalized. All similarities are cosine.

\textbf{Closed-set rank-1 identification and misID.} For each query segment, we score it against every other enrolled segment (self-comparison masked) and take the nearest-neighbor speaker. \emph{Rank-1 accuracy} is the fraction of queries whose top-scoring speaker is the true speaker (among speakers with at least one other enrolled segment). The \emph{misidentification rate} (misID) is the fraction of queries whose margin $S_{\mathrm{same}} - S_{\mathrm{diff}} < 0$, where $S_{\mathrm{same}}$ is the maximum score to the query's own speaker and $S_{\mathrm{diff}}$ the maximum to any other speaker. This is a \textbf{closed-set top-1 error} --- every query's true speaker is enrolled, and there is no reject option --- so misID = 1 $-$ rank-1 by construction; we report both only for readability. $S_{\mathrm{diff}}$ is a maximum over the full impostor \emph{segment} pool, so misID depends on gallery size $N$ and on the per-speaker segment-count distribution (see Section~\ref{sec:limitations_and_future_work}); cross-condition comparisons (Section~\ref{subsec:domain_is_genuinly_high_density}) are matched
on $N$ and the number of segments/speaker.

\textbf{$\bm{1:N}$ EER (the reject-capable operating metric).} Treating ``Is the query's top-scoring speaker the true speaker?'' as a threshold detector \emph{with a reject option}, we sweep the operating score and report the equal-error rate; genuine trials are the per-query $S_{\mathrm{same}}$ and impostor trials the per-query $S_{\mathrm{diff}}$ (paired within a query, with $S_{\mathrm{diff}}$ a maximum over the gallery, so the trials are not independent --- a nonstandard detector, distinct from a pairwise verification EER).

\textbf{Clone-probe metrics.} The gallery is every agency actor with $\geq$8 real segments (\textasciitilde1,100; the cross-file operating point restricts to the \textasciitilde900 actors with segments from $\geq$2 source files). The threshold $\tau$ is set at the real-vs-real EER operating point: genuine = held-out real segments (positions 8--12) to own centroid, impostor = to the best other centroid. Each clone of target $T$ is scored against all centroids: \emph{detection} = (score to $T \geq \tau$); \emph{NN-mismatch} = (nearest centroid $\neq T$), counted unconditionally; \emph{wrongful accusation} = (nearest $\neq T$ \textbf{and} its score $\geq$ $\tau$); \emph{collateral} = number of non-target centroids scoring $\geq$ $\tau$. Because segments are cut as consecutive windows from the same source files, genuine trials may share a source file with enrollment; we therefore also report a \emph{cross-file} operating point whose genuine trials are drawn only from source files not used in enrollment (detection rises \textasciitilde1--5\,pp). The \textbf{non-mated probe} targets the 65 out-of-gallery freelance actors (2 freelance IDs with $\geq$8 agency segments are excluded as gallery overlaps) and generates 390 clones, each with Irodori-TTS (the same 6 fixed sentences, same recipe) and Seed-VC (same recipe, resampled to 16\,kHz; all Seed-VC conversions, mated and non-mated, use the six fixed jvs001 source utterances of Section~\ref{subsec:defensive_clone_probe_and_clone_geometry}), scored against the same centered gallery at the same $\tau$; the targets' real freelance segments (positions 2--14, $n = 756$) are reported alongside as a same-channel baseline.

\textbf{Centroid $\bm{1:N}$ companion metric.} The centroid-protocol floor of Section~\ref{subsec:misidentification_floor_survives} builds one L2-normalized 8-segment centroid per eval-half speaker (\textasciitilde495 speakers, gallery-mean-centered as above), scores each held-out probe segment by cosine argmax over the gallery for the closed-set top-1 error, and for the open-set DIR@FAR holds out half the eval speakers as never-enrolled impostors, calibrating the reject threshold on the impostor top-1 score distribution (the 99th/95th percentile for a false-alarm rate of 1\,\%/5\,\%); values are mean $\pm$ std over the three speaker-disjoint splits.

\textbf{Back-ends, calibration, and cost metrics.} Discriminative back-ends are trained and evaluated on speaker-disjoint halves (55\,\% train / 45\,\% eval), averaged over three random splits ($\pm$ std). LDA is a discriminant projection reduced to $\min(150, \#\mathrm{train} - 1)$ components (kept below the embedding dim so it is distinct from WCCN, which under cosine coincides with a full-rank LDA); WCCN uses a trace-relative whitening ridge; the two-covariance PLDA-LLR reduces dimensionality by PCA to $\min(200, \#\mathrm{train} - 1)$ before estimation, uses a pooled within-class scatter and a \textbf{size-weighted between-class scatter of the class means} (a moment/scatter estimate, \emph{not} an ML/EM two-covariance fit --- the between-class scatter is an uncorrected, biased estimate of the between-class covariance parameter) with the same normalization, a trace-relative ridge, and an eigendecomposition-based $W^{-1/2}$, where $W$ is the within-class covariance. In the simultaneously diagonalized basis (within-class covariance $I$, between-class covariance $\mathrm{diag}(\lambda)$), the same/different-speaker LLR of a pair factorizes per dimension into the closed form of Section~\ref{subsec:misidentification_floor_survives}, $C + g_i + g_j + u_i^{\mathsf{T}}(q \odot u_j)$, with $q_d = \lambda_d/(2\lambda_d + 1)$, $g_i = \sum_d p_d u_{i,d}^2$, and $p_d = -\lambda_d^2/\left(2(\lambda_d + 1)(2\lambda_d + 1)\right) \leq 0$, so each $g$ is a per-vector norm penalty. As an estimator control, we additionally refit the two-covariance model by EM (the ML estimator: $x = \mu + y + e$, $y \sim \mathcal{N}(0, B)$, $e \sim \mathcal{N}(0, W)$; initialized at the moment estimates, iterated in the simultaneously diagonalized basis to convergence, $\leq$30 iterations with monotone log-likelihood), with identical preprocessing, PCA, splits, and paired sampled trials: closed-set misID changes by at most 0.4\,pp on every encoder except jxvector ($-0.9$\,pp), pairwise EER by $\leq$0.6\,pp, and $\mathrm{min}C_\mathrm{llr}$ by $\leq$0.014, so the tabulated moment-fit values carry the floor claim unchanged. ``Geometry-limited'' is therefore relative to this linear re-ranking suite, moment-fit, and EM alike. \emph{$\mathrm{min}C_\mathrm{llr}$} is the $C_\mathrm{llr}$ after optimal (PAV/isotonic) calibration; \emph{$\mathrm{act}C_\mathrm{llr}$} is the $C_\mathrm{llr}$ after an actual logistic (affine) calibration fit on half the trials and evaluated on the other half. Both are computed on sampled trials drawn with replacement from the eval speakers, so they are \textbf{in-sample optimistic} bounds (correlated trials; the $\mathrm{act}C_\mathrm{llr}$ split is not speaker-disjoint), and are reported as such --- the calibration loss $\mathrm{act}C_\mathrm{llr}$ $-$ $\mathrm{min}C_\mathrm{llr}$ is therefore a lower bound on deployed calibration loss. \emph{minDCF} (computed as a diagnostic at prior $p = 0.01/0.05$ with $C_{\mathrm{miss}} = C_{\mathrm{fa}} = 1$ for the cosine/LDA/WCCN stages, and at $p = 0.01$ for the PLDA stage) is \textbf{not tabulated} --- it is a \textbf{generic NIST/ASVspoof reference operating point}, not the asymmetric cost of the wrongful-accusation application (where $C_{\mathrm{fa}} \gg C_{\mathrm{miss}}$), and its values sit close to the reject-all baseline of 1.0 (0.45--0.98 at $p = 0.01$), so it carries no additional information beyond the $C_\mathrm{llr}$/EER already reported.

\textbf{Verification/floor trials.} For the pairwise-trial metrics, we sample a balanced 40,000 target / 40,000 impostor trial set from the evaluation speakers (targets require $\geq$2 segments; impostors are cross-speaker), with replacement. Score normalization is \textbf{AS-norm} with a top-300 impostor cohort drawn from the evaluation set --- a mild form of optimism, since the cohort is in-domain and does not test cross-condition portability. All confidence intervals are speaker-level bootstrap, resampling speakers with replacement: \textbf{300 resamples} for the AS-norm floor / critical-margin table (Table~\ref{tb:asnorm}), 2,000 for the fairness and false-match analyses, and 1,000 for the clone probe. Permutation/rewiring tests report empirical $p$-values, $p = (1 + \#\{\mathrm{null} \geq \mathrm{obs}\})/(1+B)$, floored at $1/(B+1)$; $B = 1,000$ for the confusion-graph in-degree/asymmetry nulls, $B = 200$ for modularity rewirings and the hubness null (Section~\ref{subsec:raw_geometry}).

\textbf{Style taxonomy.} Segment styles are mapped from demo-reel track names into six buckets --- narration, dialogue (\jp{セリフ}), name-read, full, freeform, and \emph{other} (46\,\% of segments, mostly generic filenames); character voices are not isolated and fall within dialogue/other (Section~\ref{subsec:cross_cutting_findings}, finding C, and Section~\ref{sec:limitations_and_future_work}).

\textbf{Confound controls.} The calibration/floor analyses are additionally run on a codec-homogeneous agency subset and with impostors restricted to different agencies; the clone channel control applies the same 7\,kHz low-pass + MP3 round-trip to real and clone audio. The vocoder copy-synthesis control analysis-resynthesizes 800 real segments (200 speakers; segment positions 8--11, which are not used for enrollment) through BigVGAN v2 (22\,kHz, 80-band mel), resamples back to 16\,kHz, and evaluates with the same protocol as the real-vs-synthetic separator (logistic regression on centered + L2-normalized embeddings, speaker-disjoint GroupKFold, balanced accuracy); the ``flagged synthetic'' rates from the real-vs-clone separator are evaluated only on the 174 speakers outside the clone-target set, whose real segments never enter that separator's training. The BGM audit scores each source file by the maximum clipwise probability of the music-related classes from an AudioSet-trained tagger (PANNs CNN14~\cite{kong2020panns}) over the first 60\,s (decision threshold 0.5, sensitivity analysis at 0.3/0.7; 300 clones serve as a music-free negative control). The content-matched control takes, for each of the 120 Irodori-TTS targets, up to 4 held-out real segments (positions 8 and later, with the reference audio excluded), transcribes them with faster-whisper large-v3, and generates speech from the same text with the identical zero-shot recipe used for the original clone generation (the same concatenated reference audio; 477 pairs, 0 synthesis failures); real-vs-content-matched-clone and real-vs-original-clone are then compared under the same separator protocol. The back-end-stage control comparison re-draws the VA subsets under the matched conditions of Section~\ref{subsec:domain_is_genuinly_high_density} (matched $N$, matched segments per speaker, shared seeds) and applies the same back-ends as Table~\ref{tb:reranking} (with the LDA and pre-PLDA PCA dimensions capped at the number of training speakers $-$ 1) over three speaker-disjoint splits.

\textbf{Bootstrap confidence intervals.} Table~\ref{tb:ci} reports the speaker-level bootstrap 95\,\% CIs behind the point estimates of Tables~\ref{tb:asnorm} and~\ref{tb:clone_probe} (computed for the misID/rank-1 quantities and clone-probe rates; the $1:N$ EER is a point estimate). Table~\ref{tb:asnorm}-derived CIs use 300 resamples; the clone probe 1,000 (Appendix~A).

\begin{table}[!t]
\centering
\caption{Speaker-level bootstrap 95\,\% CIs. Top: AS-norm-stage closed-set misID (companion to Table~\ref{tb:asnorm}). Bottom: clone-probe attribution recall and wrongful accusation (companion to Table~\ref{tb:clone_probe}).}
\label{tb:ci}
\scriptsize
\setlength{\tabcolsep}{3pt}
\begin{tabular}{lrr}
\toprule
Encoder & misID\,\% & 95\,\% CI \\
\midrule
ECAPA-TDNN & 12.60 & [11.81, 13.40] \\
CAM++ & 6.66 & [6.22, 7.19] \\
ReDimNet-b2 & 6.78 & [6.28, 7.37] \\
animeva & 2.18 & [1.92, 2.53] \\
ensemble SV-4 & 4.38 & [4.01, 4.81] \\
ensemble all-6 & 4.89 & [4.53, 5.37] \\
\bottomrule
\end{tabular}

\medskip

\begin{tabular}{llrr}
\toprule
Synthesizer & Encoder & Attr.\ recall\,\% & Wrongful acc.\,\% \\
\midrule
Seed-VC & animeva & 81.8 [76.0, 87.5] & 1.5 [0.6, 2.6] \\
Seed-VC & ECAPA-TDNN & 68.2 [61.0, 74.9] & 19.0 [13.5, 24.7] \\
Irodori-TTS & animeva & 67.5 [61.2, 73.5] & 4.7 [2.9, 6.7] \\
Irodori-TTS & ECAPA-TDNN & 51.0 [43.2, 58.2] & 17.6 [12.6, 22.5] \\
GPT-SoVITS v1 & animeva & 2.9 [0.4, 6.2] & 2.9 [0.8, 5.4] \\
GPT-SoVITS v1 & ECAPA-TDNN & 7.1 [1.2, 15.0] & 15.4 [7.5, 25.4] \\
GPT-SoVITS v2 & animeva & 14.6 [6.2, 24.6] & 10.0 [4.6, 17.1] \\
GPT-SoVITS v2 & ECAPA-TDNN & 6.2 [1.7, 12.5] & 15.8 [7.1, 26.7] \\
GPT-SoVITS v3 & animeva & 59.2 [47.1, 71.7] & 3.8 [0.8, 7.1] \\
GPT-SoVITS v3 & ECAPA-TDNN & 44.2 [30.4, 57.5] & 12.5 [5.8, 20.0] \\
GPT-SoVITS v4 & animeva & 64.2 [51.7, 76.2] & 5.4 [1.7, 9.6] \\
GPT-SoVITS v4 & ECAPA-TDNN & 51.2 [37.9, 64.2] & 14.6 [7.1, 23.3] \\
GPT-SoVITS v2ProPlus & animeva & 55.8 [42.1, 69.2] & 5.0 [2.5, 7.5] \\
GPT-SoVITS v2ProPlus & ECAPA-TDNN & 46.2 [32.1, 60.0] & 17.5 [8.3, 28.3] \\
\bottomrule
\end{tabular}
\end{table}

\section*{Data and code availability}

\textbf{Data availability statement.} Analysis code and anonymized, hash-keyed derived statistics are publicly available; the raw speaker embeddings are biometric data and are withheld, available to reviewers/replicators on request under a data-use agreement for verification only (details below). Code and derived artifacts are released in two tiers that separate what is public from what requires rights care. \emph{Public tier} (\textbf{\url{https://github.com/shuheikatoinfo/va-space-geometry}}; archived at Zenodo, concept DOI \url{https://doi.org/10.5281/zenodo.21368121}): the full analysis code --- gallery construction, hubness/misID/EER computation, the back-end suite including the EM refit, and all clone probes and confound controls --- together with the anonymized derived artifacts: per-encoder derived statistics, speaker-disjoint split definitions and random seeds, and the training-overlap hash set. The core geometry and back-end suite are data-independent and run on any embedding set conforming to the documented \texttt{output/embeddings/<model>.npz} layout (a per-segment embedding matrix plus speaker-ID, segment, style, and recording-source keys); the clone probes, demographic and session-disjoint analyses, and audio-based confound controls additionally require the withheld segment-level metadata (and, where applicable, the corresponding audio), available to reviewers/replicators on request. All per-speaker artifacts are \textbf{keyed by salted one-way hashes of the registry ID} (not the seigura ID or name), with the ID $\leftrightarrow$ hash mapping withheld and available to reviewers/replicators under a data-use agreement (Section~\ref{sec:ethics_and_reproducibility}); clone-probe results that would name an innocent wrongfully-attributed actor are released only as \textbf{aggregate rates}, not per-clone rows. \emph{Withheld tier} (not publicly released): the hash-keyed per-segment and centroid embeddings and paired trial lists are biometric identifiers (Section~\ref{sec:ethics_and_reproducibility}) and are \textbf{not} published; they are available to reviewers/replicators on request under a data-use agreement (research use only, verification only). Raw collected audio and generated clones are \textbf{not} redistributed (Section~\ref{sec:ethics_and_reproducibility}). Aggregate and centroid-level results (the clone probe, threshold analyses, and confusion structure) are provided as public hash-keyed artifacts and re-derivable from the withheld embeddings on request, and the core pipeline runs on any embedding set in the documented layout, including the consented control corpora used here (JVS, CommonVoice); the segment-level results of Tables~\ref{tb:raw_geometry}--\ref{tb:reranking} can additionally be verified against the withheld segment embeddings on request.

%% file: figures.tex

\emph{Additional diagnostic figures --- style-controlled misidentification, the GPT-SoVITS version-fidelity gradient, clone-detection rates, and 2-D embedding maps --- are provided with the released code and are not required to follow the argument.}

%% file: acknowledgment.tex
\section*{Acknowledgment}
The author, a non-native English speaker, used generative AI systems as tools throughout this work, under the author's direction. Google Gemini 3.5 Flash~\cite{ai_gemini} was used to help refine the research topic. Anthropic's Claude (Fable~5, Opus~4.8, and Sonnet~5)~\cite{ai_claude} was used to discuss and refine the study design and analysis, to draft the article's English-language text, and to generate the figures, tables, and the analysis and experiment code, including running some tasks agentically. The author directed this work in detail throughout---setting the requirements, critically examining the methodology, procedures, and outputs, and rejecting and re-directing the work until it met the author's specifications---substantially revised the manuscript text by hand, made all research and methodological decisions, and takes full responsibility for the entire content of the article.